\documentclass[twocolumn,aps,prl,floatfix,superscriptaddress,a4paper,showpacs,showkeys,nofootinbib,notitlepage]{revtex4-2}
\usepackage[colorlinks=true,linktocpage=true,linkcolor=blue,citecolor=blue,allcolors=blue]{hyperref}
\usepackage{epsfig}
\usepackage{latexsym}
\usepackage[utf8]{inputenc}
\usepackage{xspace}
\usepackage{indentfirst}
\usepackage{enumitem}
\usepackage{color}
\usepackage{placeins}

\usepackage{setspace}
\usepackage{lipsum}

\usepackage{hyperref}

\usepackage{todonotes}

\usepackage{comment}

\usepackage{amsmath}
\usepackage{amssymb}
\usepackage[english]{babel}
\usepackage{url}
\usepackage[T1]{fontenc}

\graphicspath{{figs/}}
\newcommand{\mean}[1]{\langle #1 \rangle}
\newcommand{\eq}[1]{\begin{align} #1 \end{align}}

\newcommand{\be}{\begin{equation}}
\newcommand{\ee}{\end{equation}}

\usepackage[normalem]{ulem} % for \sout

\begin{document}
\title{Probing vorticity through femtoscopic correlations}
\author{Oleh Savchuk}\thanks{Corresponding author}
\email{savchuk@frib.msu.edu} 
\affiliation{Facility for Rare Isotope Beams, Michigan State University, East Lansing, MI 48824 USA}
\author{Pawel Danielewicz}
\affiliation{Facility for Rare Isotope Beams, Michigan State University, East Lansing, MI 48824 USA}
\author{D\'aniel Kincses}
\affiliation{ELTE E\"otv\"os Lor\'and University, P\'azm\'any P\'eter s\'et\'any 1/A, Budapest, Hungary}
\author{Agnieszka Sorensen}
\affiliation{Facility for Rare Isotope Beams, Michigan State University, East Lansing, MI 48824 USA}
\date{\today}

\begin{abstract}
In heavy-ion collisions, as the two nuclei pass through one another and create hot and dense matter, part of their initial angular momentum is transferred to the fireball, generating a nonzero average vorticity.
Understanding heavy-ion collision dynamics and its influence on key observables, including those used to probe the initial state or assess thermodynamics of nuclear matter, requires understanding the magnitude of effects tied to vorticity. In this work, 
we use simulations of non-central Au+Au collisions at $E_{\rm{kin}}=1.23~A\rm{GeV}$ to show that the rotation of the system impacts the space-time picture of particle emission and, in particular, leaves imprints on proton-pion femtoscopic correlations. 
Next, we use coarse-graining of the simulation outputs to extract the collective velocity as a function of position and time, shedding light on the dynamical origin of this effect.
Moreover, we demonstrate that the displacement between the proton and pion emission centers quantifies the strength of the rotation and propose it as a new signal of vorticity in heavy-ion collisions.
\end{abstract}
\keywords{}

\maketitle

\section{Introduction}
\label{Introduction}

Collisions of heavy nuclei at relativistic velocities provide a unique means of studying nuclear matter at extreme densities and temperatures, including regions inaccessible to first-principle or \textit{ab initio} calculations.
At ultra-relativistic energies, explored at the Relativistic Heavy Ion Collider~(RHIC) and at the Large Hadron Collider~(LHC), heavy-ion collisions enable studies of the emergence of quark and gluon degrees of freedom and the dynamics of the quark-gluon plasma~\cite{STAR:2005gfr,PHENIX:2004vcz,BRAHMS:2004adc,PHOBOS:2004zne,Muller:2012zq,Elfner:2022iae}.
In the intermediate regime, characterized by the center-of-mass energies per nucleon pair $\sqrt{s_{\rm{NN}}} \sim 10~\rm{GeV}$ and explored in the RHIC Beam Energy Scan~(BES) program, heavy-ion collisions probe the phase diagram of strongly interacting matter~\cite{Odyniec:2013kna,STAR:2017sal,Bzdak:2019pkr,Du:2024wjm,Savchuk:2024ykb}.
At low energies, probed by the Fixed-Target campaign of BES, by the HADES experiment at GSI, and in future experiments at the Facility for Antiproton and Ion Research~(FAIR), heavy-ion collisions are used to infer the dense nuclear matter equation of state~(EOS).
Moreover, experiments colliding radioactive beams of neutron- and proton-rich nuclei, such as those at the Facility for Rare Isotope Beams~(FRIB), probe the isospin dependence of EOS~\cite{Sorensen:2023zkk,Brown:2024rml}, complementing advances in \textit{ab initio} methods~\cite{Lynn:2019rdt,Drischler:2021kxf}, inferences from neutron star properties~\cite{Ozel:2016oaf,Lattimer:2021emm}, and multi-messenger observations of neutron star mergers~\cite{Essick:2021kjb,Dietrich:2020efo}.

A collision of two heavy nuclei creates a region of hot, dense, and strongly-interacting matter, often called a fireball. 
Properties of this matter are inferred from observables such as electromagnetic probes~\cite{Shuryak:1978ij,McLerran:1984ay,Rapp:1999ej} which are sensitive to the early evolution, collective flow~\cite{Danielewicz:1985hn,Rischke:1995pe,Stoecker:2004qu,Brachmann:1999xt,Brachmann:1999mp,Oliinychenko:2022uvy,STAR:2017ykf} which reflects the bulk dynamics, femtoscopic correlations~\cite{Lednicky:1995vk,Lednicky:2005tb,Pratt:1984su,Wiedemann:1997cr,Khyzhniak:2024chj} revealing the spatial extent and shape of the system, and polarization~\cite{Becattini:2020ngo} which probes effects driven by the angular momentum and rotation of the fireball.

In particular, a finite angular momentum causes a deviation of the collective motion from a purely radial or cylindrical expansion. This deviation can be quantified with vorticity, \textit{i.e.}, the circulation of the velocity-field \mbox{$\omega^i=\varepsilon^{ijk}\partial_j v_k$}.
Since vorticity can be inferred from the polarization of particles such as the $\Lambda$ hyperon~\cite{Becattini:2020ngo,Florkowski:2018ahw,Palermo:2024tza}, it is a subject of numerous experimental efforts. Indeed, vorticity and its dependence on energy have received considerable attention at RHIC~\cite{STAR:2017ckg}. 
At low collision energies, such as those achieved in the HADES experiment~\cite{HADES:2022enx}, recent studies of $\Lambda$ polarization indicate an increase in vorticity in the central region of the fireball, which may have a large influence on the collective flow.

However, quantifying this influence is difficult, as the polarization of $\Lambda$s is affected by several factors beyond the vorticity, including nuclear spin alignment prior to the collision and relaxation processes during the evolution~\cite{Giacalone:2025bgm} or properties of the freeze-out hypersurface~\cite{Becattini:2020ngo, Becattini:2013fla}, leading to systematic uncertainties in measuring vorticity through $\Lambda$ polarization. 
Moreover, studying vorticity at very low energies, where the yields of $\Lambda$s approach zero, also calls for an alternative observable.
In this work, we develop such an alternative by demonstrating that vorticity impacts femtoscopic correlations and by providing quantitative measures of this effect.

\section{Femtoscopy}
\label{Femtoscopy}

During the expansion phase of a heavy-ion collision, individual particles leave the fireball and stream freely toward detectors.
Femtoscopy connects the distribution of relative distances between pairs of particles at the time of emission (\textit{i.e.}, the coordinate-space characteristics of the system) to the distribution of their relative momenta at detection. 
This connection is described by the Koonin-Pratt formula~\cite{Koonin:1977fh,Pratt:1986cc} for the pair correlation function~$C_{\vec{v}}(\vec{q})$,
\begin{equation}
C_{\vec{v}}(\vec{q})=  \int \mathrm{d}\vec{r}\, K(\vec{q},\vec{r})S_{\vec{v}}(\vec{r}) ~,
\label{eq:Koonin-Pratt}
\end{equation}
where $\vec{v}$ is the velocity of the center-of-mass of a particle pair, $\vec{q}$ is the relative momentum of particles within the pair and $\vec{r}$ is the distance between those particles at emission (both in the rest frame of the pair), $S_{\vec{v}}(\vec{r})$ is a function that describes the distribution of relative distances between particles (known as the source function), and $K(\vec{q},\vec{r})$ is the squared relative scattering wave function $|\psi(\vec{q},\vec{r})|^2$ which gives the quantum-mechanical probability for a pair separated by $\vec{r}$ to be found in a final momentum state $\vec{q}$. 
Using Eq.~\eqref{eq:Koonin-Pratt} to establish a connection between the measured pair correlations $C_{\vec{v}}(\vec{q})$ and the source function $S_{\vec{v}}(\vec{r})$ assumes that particles leaving the fireball are weakly correlated, that correlations at small relative momentum $\vec{q}$ (\textit{i.e.}, small relative velocity) do not involve particles outside the pair, that particles are emitted at similar times in the pair center-of-mass frame, and that the source $S_{\vec{v}}(\vec{r})$ is a smooth function with respect to~$\vec{q}$~\cite{Lisa:2005dd}. All of these assumptions are expected to hold reasonably well in heavy-ion collisions~\cite{Pratt:1997pw,Rzesa:2024oqp}.

In experiment, pair correlations are determined by dividing the two-particle probability distribution by a product of the single-particle distributions,
\eq{
C_{\vec{v}}(\vec{q}) =\frac{ \frac{d^6P_{12}}{dp_1^3dp_2^3} }{\frac{d^3P_1}{dp_1^3}\frac{d^3P_2}{dp_2^3}}~,
}
where indices $1$ and $2$ correspond to particles in a pair, $\vec{p}_i$ is the measured momentum, $\vec{v}=\frac{\vec{p}_1+\vec{p}_2}{E_1+E_2}$ is the center-of-mass velocity of the pair, $\vec{q}=(\vec{p}_1-\vec{p}_2) +(\gamma-1)\big[(\vec{p}_1-\vec{p}_2)\cdot \vec{v}\big]\frac{\vec{v}}{|\vec{v}|^2} - \gamma(E_1-E_2)\vec{v}$ is the relative momentum of the particles in the pair in their center-of-mass frame, and $\gamma=(1-\vec{v}^2)^{-\frac{1}{2}}$.
If the kernel $K(\vec{q},\vec{r})$ is known, then one can extract the source function
$S_{\vec{v}}(\vec{r})$ by inverting Eq.~\eqref{eq:Koonin-Pratt}. 
Alternatively, one can compare experimental measurements of $C_{\vec{v}}(q)$ with theoretical results obtained through Eq.~\eqref{eq:Koonin-Pratt} for a given source function~$S_{\vec{v}}(\vec{r})$%; in particular, dynamical simulations can be used to obtain $S_{\vec{v}}(\vec{r})$ for the system of interest
.

\begin{figure}
\includegraphics[width=.49\textwidth]{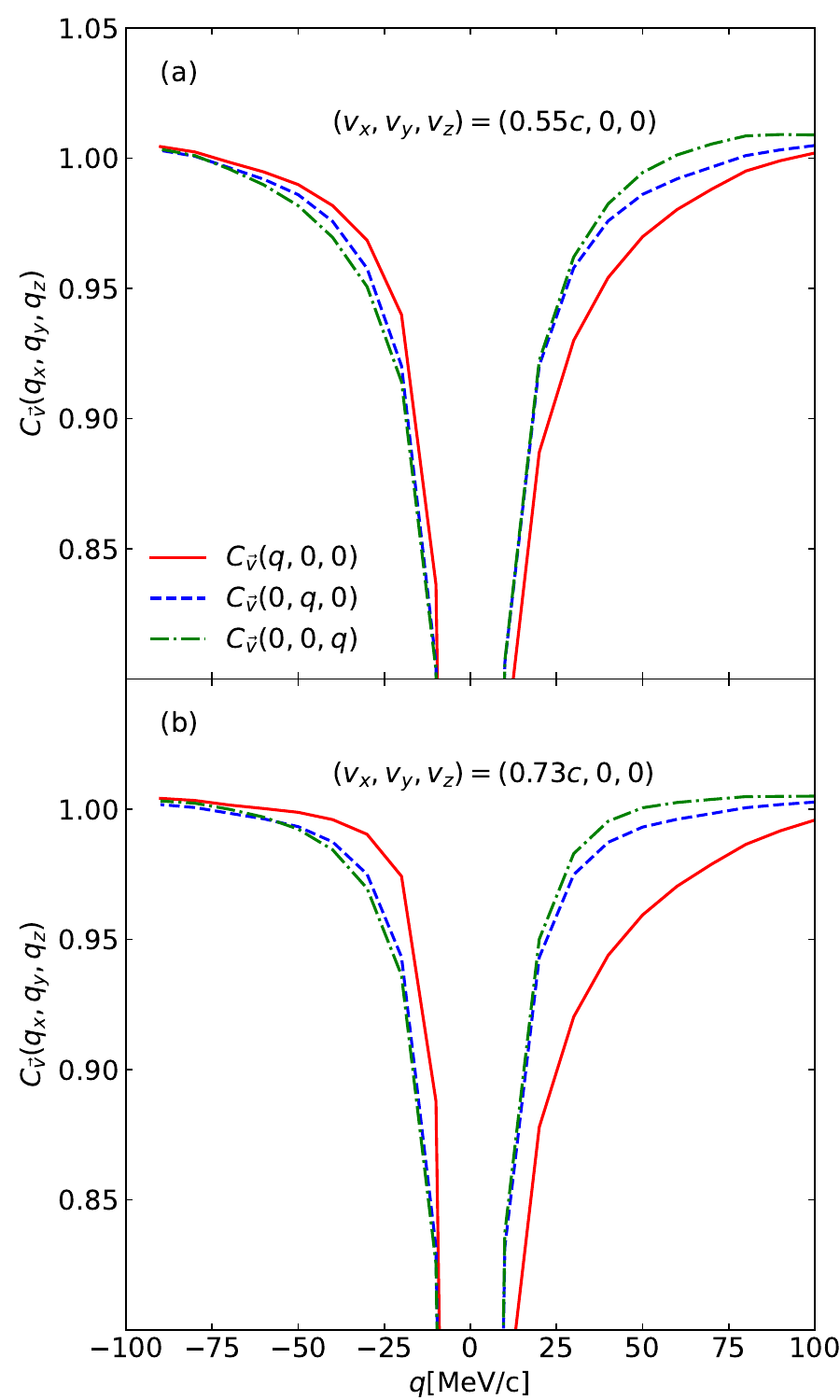}
\vspace{-5mm} % reduce white space
\caption{\label{Cq}
Correlation functions of $p\pi^+$ pairs with $\vec{v}_{p \pi} \approx v_{p \pi}^x \vec{e}_x$, for two values of $v^x_{p \pi}$ (top and bottom panel), obtained \textit{via} Eq.~\eqref{eq:Koonin-Pratt} using source functions from \texttt{UrQMD} simulations and the \texttt{Coral} kernel. 
The solid red line shows correlations along the \( q_x \) relative momentum axis, \( C_{\vec{v}}(q,0,0) \), while the dashed blue and dash-dotted green lines correspond to correlations along the \( q_y \) and \( q_z \) axes, respectively.  
For \( C_{\vec{v}}(q,0,0) \), the asymmetry around \( q = 0 \) stems from a shift of the $p\pi^+$ source toward positive values of~$x$, while the asymmetry in \( C_{\vec{v}}(0,0,q) \) reflects a smaller negative shift in the \( z \)-direction.
}
\end{figure}

In this paper, we focus on proton--pion correlations. Previously, significant effort has been dedicated to proton-proton~\cite{STAR:2005rpl, Stefaniak:2024fkf,Stefaniak:2024eux} and pion-pion~\cite{STAR:2014shf,ALICE:2011dyt,HADES:2018gop} correlations. Analyses of correlations between nonidentical particles, such as $\pi p$, $\pi K$, or $pK$, have also been performed \cite{Voloshin:1997jh,zawisza2011meson,Zawisza:2010az,Poniatowska:2015zca,ALICE:2020mkb,Wang:2024ykl}, but generally received less attention. The source function~$S_{\vec{v}}(\vec{r})$ for nonidentical particles has several distinct features compared to the identical-particle case. First, quantum statistical effects due to Bose–Einstein or Fermi–Dirac correlations are absent, so that the correlation kernel reflects only strong and Coulomb interactions. Second, $S_{\vec{v}}(\vec{r})$ is not necessarily an even function of $\vec{r}$ and thus may exhibit a finite dipole moment, $\mean{\vec{r}}\neq 0$. The lack of reflection symmetry in source functions of nonidentical particles generally results in similarly asymmetric correlations, $C_{\vec{v}}(\vec{q}) \neq C_{\vec{v}}(-\vec{q})$~\cite{Brown:1997ku}.
We use these features of proton-pion femtoscopy to study vorticity.

We extracted source functions~$S_{\vec{v}}(\vec{r})$ from simulations using the \texttt{UrQMD} transport model~\cite{Bass:1998ca,Bleicher:1999xi,Bleicher:2022kcu}, which describes A+A collisions 
in terms of explicit phase-space propagation of hadrons combined with their elastic and inelastic two-body reactions and decays of unstable particles. 
The interactions among baryons are implemented \textit{via} a potential energy per baryon dependent on
baryon density $n_B$, which allows one to incorporate any \mbox{$n_B$-dependent}
EOS in the non-relativistic Hamilton equations of motion~\cite{OmanaKuttan:2022the,Steinheimer:2022gqb,Savchuk:2022msa}.
In the present work, we used an EOS derived from the Chiral SU(3)-flavor parity-doublet Polyakov-loop quark-hadron mean-field model (CMF) \cite{Steinheimer:2010ib, Steinheimer:2011ea, Mukherjee:2016nhb, Motornenko:2018hjw} in its
most recent version~\cite{Motornenko:2019arp}. 
The CMF model provides a realistic description of nuclear matter with a nuclear incompressibility of \mbox{$K_0=267~\rm{MeV}$}, chiral symmetry breaking in the hadronic and quark sectors, and an effective deconfinement transition. 
To study the bulk evolution of matter, we simulated $4\times10^6$ collision events for midcentral ($b=6.6$ fm, or $15$-$25\%$ centrality bin~\cite{HADES:2017def}) Au+Au collisions at $E_{\rm kin}=1.23~A\mathrm{GeV}$. 
In this analysis, we considered the proton--pion pair velocities $\vec{v}_{p \pi}$ at midrapidity in the reaction plane, that is, $\vec{v}_{p \pi}\approx v_{p \pi}^x\vec{e}_x$ in the center-of-mass frame of the projectile and target, and the source functions are calculated at $t=50~\mathrm{fm/c}$.

To calculate $C_{\vec{v}}(\vec{q})$ from the source function \textit{via} Eq.~\eqref{eq:Koonin-Pratt}, we evaluated the kernel $K(\vec{q}, \vec{r})$, which encodes the two-particle interaction of a $p\pi^+$ pair, using the \texttt{Coral} package~\cite{coral}. 
This framework incorporates experimentally measured phase shifts to capture the final-state interactions between nonidentical particles, including protons and pions, as well as electromagnetic Coulomb forces. 

The correlation functions obtained are shown in Fig.~\ref{Cq}. 
The overall shape of $C_{\vec{v}}(\vec{q})$ is dictated by Coulomb repulsion, which yields a strong suppression (anti-correlation) at small relative momenta (\mbox{$|\vec{q}| \lessapprox 50~\mathrm{MeV/c}$}), and by $\Delta^{++}$-resonance scattering \textit{via} the strong interaction, leading to attraction (correlation) in the $p$-wave channel at $|\vec{q}| \gtrapprox  100~\mathrm{MeV/c}$.
The symmetric shape of the correlation along the Cartesian \( q_y \) momentum axis, $C_{\vec{v}}(0, q, 0)$, reflects the mirror symmetry of A+A heavy-ion collision systems with respect to the \( y = 0 \) plane. In contrast, the correlation functions evaluated along the \( q_x \) and \( q_z \) momentum axes, $C_{\vec{v}}(q, 0, 0)$ and $C_{\vec{v}}(0, 0, q)$, are asymmetric around the origin. 
This asymmetry corresponds to a positive shift in $\langle x_p - x_\pi \rangle$ and a negative shift in $\langle z_p - z_\pi \rangle$~\cite{Brown:1997ku,Voloshin:1997jh}. From the magnitude of the asymmetries, we can infer that the shift in the \( x \)-direction is stronger than in the \( z \)-direction. 
Below, we show that these features originate in non-zero vorticity.

\section{Source Function Analysis}
\label{connection}

\begin{figure}[t]
\vspace{-2mm}
\includegraphics[width=.49\textwidth]{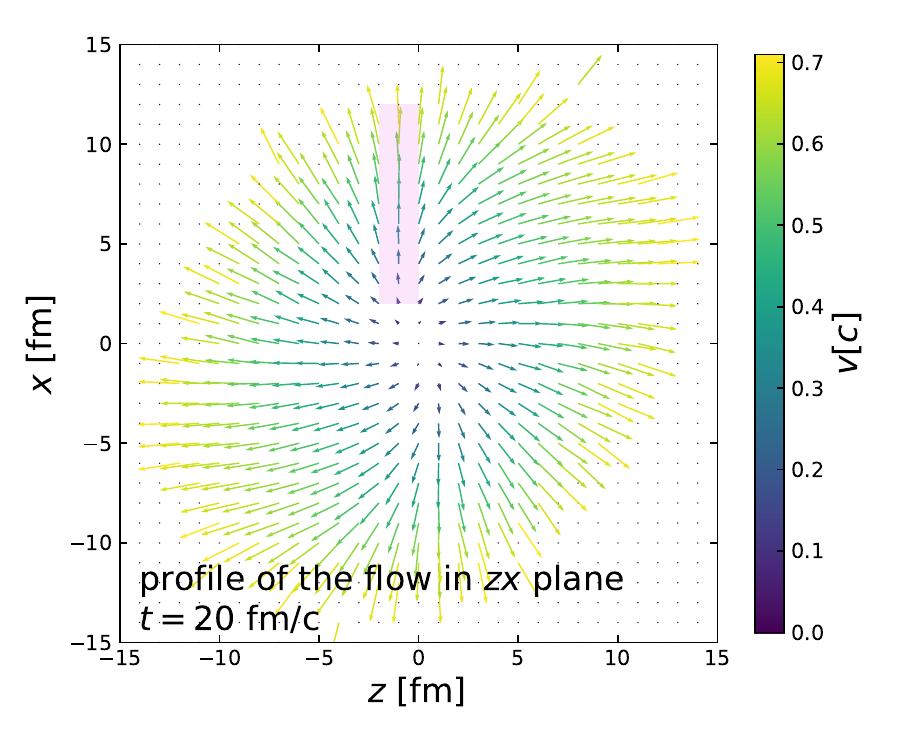}
\vspace{-10mm} % reduce white space
\caption{\label{flow-profile}
Flow in the $zx$-plane at $t = 20$~fm/c, with arrows depicting the magnitude and direction of the collective flow velocity~$\vec{v}_f$; the magnitude of $\vec{v}_f$ is also indicated by the color of the arrows (see legend). 
In our coordinate setup, the nucleus with a center of mass located at \( x > 0 \) carries a positive \( z \)-momentum, while the one at $x<0$ carries a negative \( z \)-momentum.
The expansion accelerates with distance from the center of the collision. 
The presence of angular momentum in the flow, resulting in a non-zero average vorticity, can be seen in components of $\vec{v}_f$ orthogonal to lines connecting a given point to the center of the collision. In particular, the region where $\vec{v}_f$ is largely aligned with the positive $x$-axis, marked with a violet-shaded area, is located away from $z = 0$. 
}
\end{figure}

To assess the significance of the results shown in Fig.~\ref{Cq} for vorticity studies, we studied the flow generated in collisions. 
We extracted the collective flow velocity $\vec{v}_f$ by coarse-graining~\cite{Savchuk:2022aev} the \texttt{UrQMD} outputs at $t = 20$~fm/c (an average freeze-out time~\cite{Reichert:2023eev}); since the low-density and low-temperature parts of the fireball are subject to significant noise, we only considered warm $\left( T > 50~\mathrm{MeV} \right)$ and dense $\left( n > 0.16~\mathrm{fm^{-3}} \right)$ regions of the system.  
As seen in Fig.~\ref{flow-profile}, the collective expansion becomes stronger as one moves away from the center of the system, approximating a linear (Hubble-like) dependence on the radius. Notably, because of the finite value of angular momentum, there is a significant deviation of the flow from a purely radial expansion, reflected in the fact that the region of the collision with $\vec{v}_f$ parallel to the $x$-axis is located approximately $1~\mathrm{fm}$ to the left of the $z = 0~\mathrm{fm}$ line (see the violet-shaded area in Fig.~\ref{flow-profile}). 
This can be understood by considering the evolution of the system: 
After the projectile and target media come into contact at a finite impact parameter $b$, the finite angular momentum in non-central collisions modifies the collective flow field \( \vec{v}_f(t, \vec{x}) \), adding a tangential component \( v_\phi = \vec{v} \cdot \vec{e_\phi} \) in the $zx$-plane%, around the $y$-axis, 
to the radial flow \( v_{\rho} = \vec{v} \cdot \vec{e_\rho} \). The resulting non-zero vorticity -- or circulation of the $\vec{v}_f$ field -- then spreads throughout the participant zone and away from the initial contact surface~\cite{Danielewicz:1994nb}, resulting in a rotation around the \( y \)-axis.

\begin{figure}[t]
\includegraphics[width=.48\textwidth]{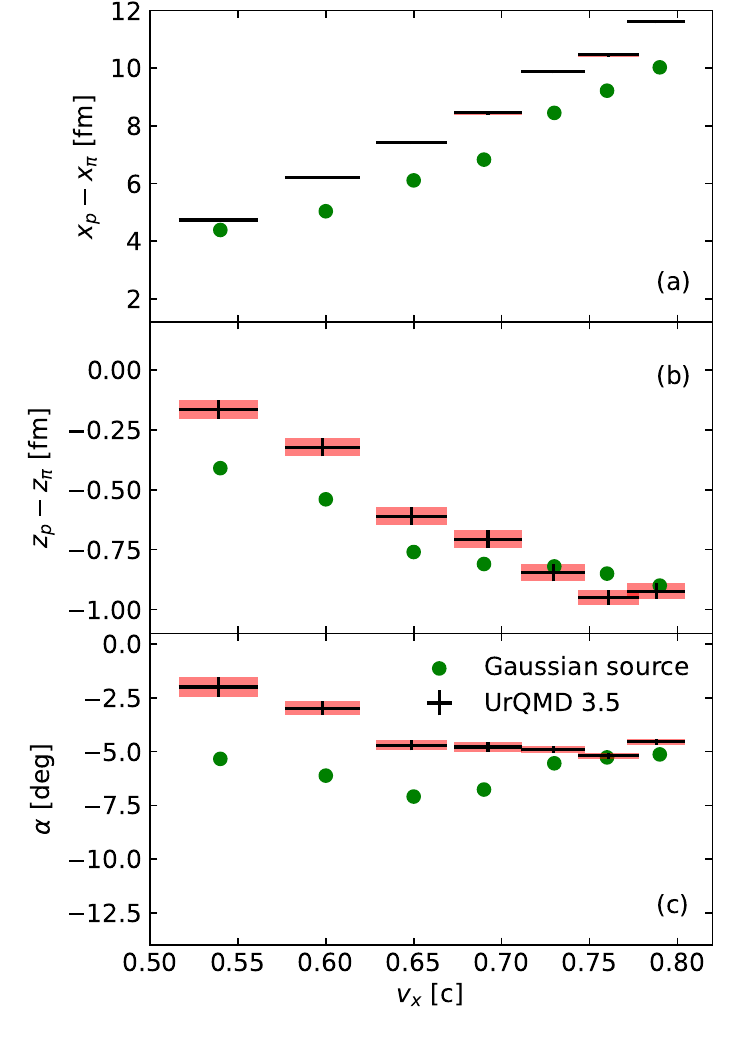}
\vspace{-10mm} % reduce white space
\caption{\label{results}
Average proton-pion source characteristics plotted against %the $x$-component of the pair velocity 
$v^x_{p \pi}$, obtained by directly averaging proton-pion pair displacements from \texttt{UrQMD} simulation outputs (black crosses with red areas denoting errors) and by extracting the source size from $p \pi^+$ femtoscopic correlations~$C_{\vec{v}}(\vec{q})$, Fig.~\ref{Cq} (green dots). Panels~(a) and (b) show the average proton-pion source displacement in the $x$- and $z$-directions, respectively. Panel~(c) shows the angle $\alpha = \angle\big( \vec{v}, \mean{\vec{r}_p - \vec{r}_{\pi}}\big)$ which can be used as a measure of  vorticity. The differences between source characteristics extracted directly and from a fit to~$C_{\vec{v}}(\vec{q})$ can be linked to assuming a Gaussian source shape in the fit.
}
\end{figure}

We investigated the influence of this non-radial flow profile on the properties of $S_{\vec{v}}(\vec{r})$. The first moment of the %proton-pion 
source is equal to the pair's average relative position,
\eq{
\int d^3r \,\vec{r}\, S_{\vec{v}}(\vec{r}) \equiv \mean{\vec{r}_p-\vec{r}_{\pi}}~.
}
We computed $\mean{\vec{r}_p-\vec{r}_{\pi}}$ for \mbox{seven center-of-mass pair} velocities, \( v^x_{p \pi}/c \in \{0.55, 0.60,  0.65, 0.69, 0.73, 0.76,0.78\}\) with $|v^y_{p \pi}/c| < 0.05$ and \mbox{$|v^z_{p \pi}/c| <0.05$}.  
This range of $v^x_{p \pi}$ reflects experimental acceptance: pions with \( p_T < 50~\mathrm{MeV/c} \) are typically undetectable, limiting the sensitivity to pion velocities \( v^x_{\pi} > 0.4c \), while very high pair velocities suffer from limited proton statistics for \( p_T > 1.0~\mathrm{GeV/c} \). The intermediate velocities explored here are optimal, especially given that they probe earlier states of the system (which have higher vorticity) while minimizing contamination from spectators and decays.

Fig.~\ref{results} shows the average proton-pion displacements in the $x$ and $z$ directions as functions of the pair velocity $v_{p \pi}^x$. 
In panel~(a), \( \mean{ x_p - x_\pi} \) is positive and increases with~$v_{p \pi}^x$, reflecting the fact that protons closely follow the collective velocity, characterized by the Hubble-like radial collective expansion seen in Fig.~\ref{flow-profile}, while pions are less sensitive to flow and are emitted almost uniformly throughout the system. 
Consequently, for higher $v_{p \pi}^x$, protons leave the system farther out from the center of the collision, while pions remain centered near the origin, \textit{i.e.}, ``behind'' protons. 
The negative values of \( \mean{z_p - z_\pi } \), seen in panel (b), reflect a non-radial (tangential) flow component driven by the system’s angular momentum. Since protons are strongly correlated with flow, they tend to originate at \( z_p < 0 \) (see the violet-shaded area in Fig.~\ref{flow-profile}), while pions again remain near the geometric center, leading to a negative \( \mean{z_p - z_\pi }\approx \mean{z_p} \). 
This behavior is consistent with the asymmetry of the $C_{\vec{v}}(q, 0, 0)$ and $C_{\vec{v}}(0, 0, q)$ correlation functions shown in Fig.~\ref{Cq} and with the coarse-grained flow profile shown in Fig.~\ref{flow-profile}.

The impact of vorticity in the fireball on the collective expansion can be quantified with the angle \( \alpha \) between the pair velocity \( \vec{v}_{p \pi} \) and the average pair separation \mbox{\( \mean{\vec{r}_p-\vec{r}_{\pi}} \)}, which is negative (positive) for negative (positive) vorticity (\textit{i.e.}, clockwise (counter-clockwise) rotation around the $y$ axis) and which can be extracted from 
\begin{equation}\label{alpha}
\sin \alpha = \frac{ |\mean{\vec{r}_p-\vec{r}_{\pi}}\times \vec{v}_{p \pi} |}{|\mean{\vec{r}_p-\vec{r}_{\pi}}||\vec{v}_{p \pi}|}~.
\end{equation}
Note that for purely radial flow, \( \sin \alpha = 0 \), while for purely rotational flow, \( \sin \alpha = \pm 1 \) (i.e., \( \alpha = \pm 90^\circ \)). 

Panel~(c) in Fig.~\ref{results} shows $\alpha$ computed from the proton-pion source according to Eq.~\eqref{alpha}; we note that in our case, where $\vec{v}_{p \pi} = v^x_{p \pi} \vec{e}_x$ and by symmetry $\mean{y_p - y_{\pi}} = 0$, only the $y$-component of the cross product is non-zero, so that $\sin \alpha =\mean{z_p - z_{\pi}}/ \sqrt{\mean{x_p - x_{\pi}}^2+\mean{z_p - z_{\pi}}^2}$.  
As expected, $\alpha$ is negative, corresponding to a clockwise rotation (when looking against the $y$-axis).
The small values of $\alpha$ indicate that the non-radial flow is much weaker than the radial flow.
The increase in vorticity with $v^x_{p \pi}$ (or, equivalently, distance from the origin) implies a non-rigid rotation of the fireball, with rotation becoming stronger away from the origin. This trend can also be observed in Fig.~\ref{flow-profile}, and it agrees with the intuition that particles emitted at higher velocities typically leave the system earlier in the evolution when the system's vorticity is stronger.
Furthermore, a similar behavior was observed in simulations using the transport code \texttt{SMASH}~\cite{SMASH:2016zqf}, indicating the robustness of our results across simulation frameworks.

\begin{figure}[t]
\includegraphics[width=.49\textwidth]{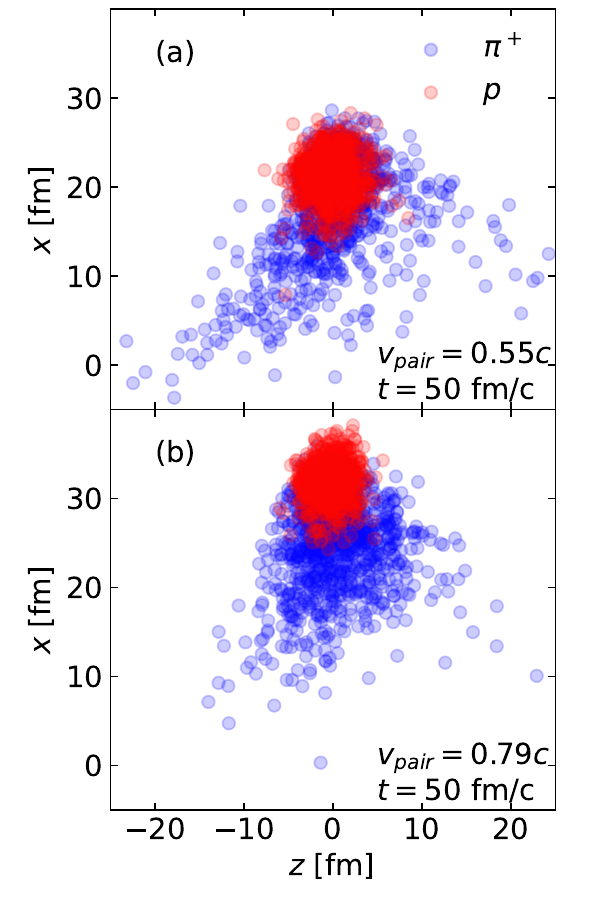}
\vspace{-9mm} % reduce white space
\caption{
Proton (red) and pion (blue) emission points, extracted from \texttt{UrQMD} simulations, for $v_{p \pi}^x = 0.55c$ (\textit{top}) and $0.73c$ (\textit{bottom}). The proton source is localized and approximates a Gaussian, while the pion source differs from a simple Gaussian, with more particles emitted from a tail at a negative $z$.  
The pion source tails are seen to be reduced for larger~$v_{p \pi}^x$.
}
\label{fig:proton_pion_source}
\end{figure}

The potential of these findings lies in the fact that moments of the source can also be extracted from experimentally measurable $p\pi^+$ femtoscopic correlations.
In fact, in Fig.~\ref{results} we also show $ \mean{ x_p - x_\pi}$, $ \mean{ z_p - z_\pi}$, and $\alpha$ as extracted from the femtoscopic correlations $p \pi^+$ shown in Fig.~\ref{Cq}.
We note that the systematic differences between the source characteristics obtained directly from the simulation output and from a fit to the computed femtoscopic correlations are likely a consequence of extracting the source size using a Gaussian \textit{ansatz} for the source functions~\cite{Savchuk:2025prep}.
In Fig.~\ref{fig:proton_pion_source}, we show the proton and pion emission points, extracted from the simulations, which indeed reveal that while the proton source is approximately Gaussian, the pion source is characterized by long tails.
Moreover, at higher velocities $v_{p\pi}^x$, where the Gaussian source assumption can be seen to be more applicable, the results of the two analyses in Fig.~\ref{results} converge.
This underscores both the necessity of using an appropriate \textit{ansatz} when extracting the source size from experimental measurements~\cite{Nzabahimana:2025ivc}
as well as the potential of such state-of-the-art measurements to shed light on vorticity in heavy-ion collision systems. We will explore such improved analyses in future work.

\section{Conclusions}
\label{Conclusions}

In this paper, we present the possibility of quantifying the vorticity of systems created in heavy-ion collisions by analyzing the proton–pion source through femtoscopic correlations. 
We show that when nonradial flow is present, a shift, orthogonal to the pair velocity~$\vec{v}_{p \pi}$, develops between the average positions of protons and pions. 
In particular, we find a correlation between the strength of the fireball rotation and the angle between $\vec{v}_{p \pi}$ and $\mean{\vec{r}_p-\vec{r}_\pi}$.
As a result, we propose proton-pion source measurements as a promising new probe of vorticity in heavy-ion collisions. 
We show that the success of the method depends on using an appropriate \textit{ansatz} for the source function when extracting the source size from femtoscopic correlations.
This can be achieved with, \textit{e.g.}, a decomposition in spherical harmonics~\cite{Danielewicz:2006hi}. 

These findings open several compelling avenues for future research. 
Unlike \(\Lambda\) polarization measurements, our method is independent of spin-related effects, which makes it potentially useful for disentangling different sources of $\Lambda$ polarization (such as shear and thermal vorticity) in hydrodynamic models of heavy-ion collisions. 
Another advantage of this method is the possibility to use it at lower energies, e.g., at FRIB, where $\Lambda$ baryons are not produced, or in asymmetric systems (e.g., $p$+$A$ collisions)~\cite{Savchuk:2025prep2} where flow may form smoke-ring structures with vanishing global polarization~\cite{Lisa:2021zkj}.

\section{Acknowledgments}

The authors thank Scott Pratt, Maria Stefaniak, and Yevheniia Khyzniak for discussions. This work was supported by the Office of Science of the Department of Energy through grants DE-FG02-03ER41259 and DE-SC0019209. 

\section{Author contributions}

O.S.\ conceived the idea, performed simulations with \texttt{UrQMD}, and wrote the initial manuscript. 
A.S.\ performed complementary simulations with \texttt{SMASH} and edited the manuscript. 
P.D.\ contributed expertise in femtoscopy and imaging and edited the manuscript. 
D.K.\ contributed expertise in experimental design and edited the manuscript. 
The authors thank Scott Pratt for suggesting the investigation of vorticity.

\bibliography{main, noninspire}

\begin{thebibliography}{81}%
\makeatletter
\providecommand \@ifxundefined [1]{%
 \@ifx{#1\undefined}
}%
\providecommand \@ifnum [1]{%
 \ifnum #1\expandafter \@firstoftwo
 \else \expandafter \@secondoftwo
 \fi
}%
\providecommand \@ifx [1]{%
 \ifx #1\expandafter \@firstoftwo
 \else \expandafter \@secondoftwo
 \fi
}%
\providecommand \natexlab [1]{#1}%
\providecommand \enquote  [1]{``#1''}%
\providecommand \bibnamefont  [1]{#1}%
\providecommand \bibfnamefont [1]{#1}%
\providecommand \citenamefont [1]{#1}%
\providecommand \href@noop [0]{\@secondoftwo}%
\providecommand \href [0]{\begingroup \@sanitize@url \@href}%
\providecommand \@href[1]{\@@startlink{#1}\@@href}%
\providecommand \@@href[1]{\endgroup#1\@@endlink}%
\providecommand \@sanitize@url [0]{\catcode `\\12\catcode `\$12\catcode `\&12\catcode `\#12\catcode `\^12\catcode `\_12\catcode `\%12\relax}%
\providecommand \@@startlink[1]{}%
\providecommand \@@endlink[0]{}%
\providecommand \url  [0]{\begingroup\@sanitize@url \@url }%
\providecommand \@url [1]{\endgroup\@href {#1}{\urlprefix }}%
\providecommand \urlprefix  [0]{URL }%
\providecommand \Eprint [0]{\href }%
\providecommand \doibase [0]{https://doi.org/}%
\providecommand \selectlanguage [0]{\@gobble}%
\providecommand \bibinfo  [0]{\@secondoftwo}%
\providecommand \bibfield  [0]{\@secondoftwo}%
\providecommand \translation [1]{[#1]}%
\providecommand \BibitemOpen [0]{}%
\providecommand \bibitemStop [0]{}%
\providecommand \bibitemNoStop [0]{.\EOS\space}%
\providecommand \EOS [0]{\spacefactor3000\relax}%
\providecommand \BibitemShut  [1]{\csname bibitem#1\endcsname}%
\let\auto@bib@innerbib\@empty
%</preamble>
\bibitem [{\citenamefont {Adams}\ \emph {et~al.}(2005)\citenamefont {Adams} \emph {et~al.}}]{STAR:2005gfr}%
  \BibitemOpen
  \bibfield  {author} {\bibinfo {author} {\bibfnamefont {J.}~\bibnamefont {Adams}} \emph {et~al.} (\bibinfo {collaboration} {STAR}),\ }\bibfield  {title} {\bibinfo {title} {{Experimental and theoretical challenges in the search for the quark gluon plasma: The STAR Collaboration's critical assessment of the evidence from RHIC collisions}},\ }\href {https://doi.org/10.1016/j.nuclphysa.2005.03.085} {\bibfield  {journal} {\bibinfo  {journal} {Nucl. Phys. A}\ }\textbf {\bibinfo {volume} {757}},\ \bibinfo {pages} {102} (\bibinfo {year} {2005})},\ \Eprint {https://arxiv.org/abs/nucl-ex/0501009} {arXiv:nucl-ex/0501009} \BibitemShut {NoStop}%
\bibitem [{\citenamefont {Adcox}\ \emph {et~al.}(2005)\citenamefont {Adcox} \emph {et~al.}}]{PHENIX:2004vcz}%
  \BibitemOpen
  \bibfield  {author} {\bibinfo {author} {\bibfnamefont {K.}~\bibnamefont {Adcox}} \emph {et~al.} (\bibinfo {collaboration} {PHENIX}),\ }\bibfield  {title} {\bibinfo {title} {{Formation of dense partonic matter in relativistic nucleus-nucleus collisions at RHIC: Experimental evaluation by the PHENIX collaboration}},\ }\href {https://doi.org/10.1016/j.nuclphysa.2005.03.086} {\bibfield  {journal} {\bibinfo  {journal} {Nucl. Phys. A}\ }\textbf {\bibinfo {volume} {757}},\ \bibinfo {pages} {184} (\bibinfo {year} {2005})},\ \Eprint {https://arxiv.org/abs/nucl-ex/0410003} {arXiv:nucl-ex/0410003} \BibitemShut {NoStop}%
\bibitem [{\citenamefont {Arsene}\ \emph {et~al.}(2005)\citenamefont {Arsene} \emph {et~al.}}]{BRAHMS:2004adc}%
  \BibitemOpen
  \bibfield  {author} {\bibinfo {author} {\bibfnamefont {I.}~\bibnamefont {Arsene}} \emph {et~al.} (\bibinfo {collaboration} {BRAHMS}),\ }\bibfield  {title} {\bibinfo {title} {{Quark gluon plasma and color glass condensate at RHIC? The Perspective from the BRAHMS experiment}},\ }\href {https://doi.org/10.1016/j.nuclphysa.2005.02.130} {\bibfield  {journal} {\bibinfo  {journal} {Nucl. Phys. A}\ }\textbf {\bibinfo {volume} {757}},\ \bibinfo {pages} {1} (\bibinfo {year} {2005})},\ \Eprint {https://arxiv.org/abs/nucl-ex/0410020} {arXiv:nucl-ex/0410020} \BibitemShut {NoStop}%
\bibitem [{\citenamefont {Back}\ \emph {et~al.}(2005)\citenamefont {Back} \emph {et~al.}}]{PHOBOS:2004zne}%
  \BibitemOpen
  \bibfield  {author} {\bibinfo {author} {\bibfnamefont {B.~B.}\ \bibnamefont {Back}} \emph {et~al.} (\bibinfo {collaboration} {PHOBOS}),\ }\bibfield  {title} {\bibinfo {title} {{The PHOBOS perspective on discoveries at RHIC}},\ }\href {https://doi.org/10.1016/j.nuclphysa.2005.03.084} {\bibfield  {journal} {\bibinfo  {journal} {Nucl. Phys. A}\ }\textbf {\bibinfo {volume} {757}},\ \bibinfo {pages} {28} (\bibinfo {year} {2005})},\ \Eprint {https://arxiv.org/abs/nucl-ex/0410022} {arXiv:nucl-ex/0410022} \BibitemShut {NoStop}%
\bibitem [{\citenamefont {Muller}\ \emph {et~al.}(2012)\citenamefont {Muller}, \citenamefont {Schukraft},\ and\ \citenamefont {Wyslouch}}]{Muller:2012zq}%
  \BibitemOpen
  \bibfield  {author} {\bibinfo {author} {\bibfnamefont {B.}~\bibnamefont {Muller}}, \bibinfo {author} {\bibfnamefont {J.}~\bibnamefont {Schukraft}},\ and\ \bibinfo {author} {\bibfnamefont {B.}~\bibnamefont {Wyslouch}},\ }\bibfield  {title} {\bibinfo {title} {{First Results from Pb+Pb collisions at the LHC}},\ }\href {https://doi.org/10.1146/annurev-nucl-102711-094910} {\bibfield  {journal} {\bibinfo  {journal} {Ann. Rev. Nucl. Part. Sci.}\ }\textbf {\bibinfo {volume} {62}},\ \bibinfo {pages} {361} (\bibinfo {year} {2012})},\ \Eprint {https://arxiv.org/abs/1202.3233} {arXiv:1202.3233 [hep-ex]} \BibitemShut {NoStop}%
\bibitem [{\citenamefont {Elfner}\ and\ \citenamefont {M{\"u}ller}(2023)}]{Elfner:2022iae}%
  \BibitemOpen
  \bibfield  {author} {\bibinfo {author} {\bibfnamefont {H.}~\bibnamefont {Elfner}}\ and\ \bibinfo {author} {\bibfnamefont {B.}~\bibnamefont {M{\"u}ller}},\ }\bibfield  {title} {\bibinfo {title} {{The exploration of hot and dense nuclear matter: introduction to relativistic heavy-ion physics}},\ }\href {https://doi.org/10.1088/1361-6471/ace824} {\bibfield  {journal} {\bibinfo  {journal} {J. Phys. G}\ }\textbf {\bibinfo {volume} {50}},\ \bibinfo {pages} {103001} (\bibinfo {year} {2023})},\ \Eprint {https://arxiv.org/abs/2210.12056} {arXiv:2210.12056 [nucl-th]} \BibitemShut {NoStop}%
\bibitem [{\citenamefont {Odyniec}(2013)}]{Odyniec:2013kna}%
  \BibitemOpen
  \bibfield  {author} {\bibinfo {author} {\bibfnamefont {G.}~\bibnamefont {Odyniec}},\ }\bibfield  {title} {\bibinfo {title} {{The RHIC Beam Energy Scan program in STAR and what's next ...}},\ }\href {https://doi.org/10.1088/1742-6596/455/1/012037} {\bibfield  {journal} {\bibinfo  {journal} {J. Phys. Conf. Ser.}\ }\textbf {\bibinfo {volume} {455}},\ \bibinfo {pages} {012037} (\bibinfo {year} {2013})}\BibitemShut {NoStop}%
\bibitem [{\citenamefont {Adamczyk}\ \emph {et~al.}(2017{\natexlab{a}})\citenamefont {Adamczyk} \emph {et~al.}}]{STAR:2017sal}%
  \BibitemOpen
  \bibfield  {author} {\bibinfo {author} {\bibfnamefont {L.}~\bibnamefont {Adamczyk}} \emph {et~al.} (\bibinfo {collaboration} {STAR}),\ }\bibfield  {title} {\bibinfo {title} {{Bulk Properties of the Medium Produced in Relativistic Heavy-Ion Collisions from the Beam Energy Scan Program}},\ }\href {https://doi.org/10.1103/PhysRevC.96.044904} {\bibfield  {journal} {\bibinfo  {journal} {Phys. Rev. C}\ }\textbf {\bibinfo {volume} {96}},\ \bibinfo {pages} {044904} (\bibinfo {year} {2017}{\natexlab{a}})},\ \Eprint {https://arxiv.org/abs/1701.07065} {arXiv:1701.07065 [nucl-ex]} \BibitemShut {NoStop}%
\bibitem [{\citenamefont {Bzdak}\ \emph {et~al.}(2020)\citenamefont {Bzdak}, \citenamefont {Esumi}, \citenamefont {Koch}, \citenamefont {Liao}, \citenamefont {Stephanov},\ and\ \citenamefont {Xu}}]{Bzdak:2019pkr}%
  \BibitemOpen
  \bibfield  {author} {\bibinfo {author} {\bibfnamefont {A.}~\bibnamefont {Bzdak}}, \bibinfo {author} {\bibfnamefont {S.}~\bibnamefont {Esumi}}, \bibinfo {author} {\bibfnamefont {V.}~\bibnamefont {Koch}}, \bibinfo {author} {\bibfnamefont {J.}~\bibnamefont {Liao}}, \bibinfo {author} {\bibfnamefont {M.}~\bibnamefont {Stephanov}},\ and\ \bibinfo {author} {\bibfnamefont {N.}~\bibnamefont {Xu}},\ }\bibfield  {title} {\bibinfo {title} {{Mapping the Phases of Quantum Chromodynamics with Beam Energy Scan}},\ }\href {https://doi.org/10.1016/j.physrep.2020.01.005} {\bibfield  {journal} {\bibinfo  {journal} {Phys. Rept.}\ }\textbf {\bibinfo {volume} {853}},\ \bibinfo {pages} {1} (\bibinfo {year} {2020})},\ \Eprint {https://arxiv.org/abs/1906.00936} {arXiv:1906.00936 [nucl-th]} \BibitemShut {NoStop}%
\bibitem [{\citenamefont {Du}\ \emph {et~al.}(2024)\citenamefont {Du}, \citenamefont {Sorensen},\ and\ \citenamefont {Stephanov}}]{Du:2024wjm}%
  \BibitemOpen
  \bibfield  {author} {\bibinfo {author} {\bibfnamefont {L.}~\bibnamefont {Du}}, \bibinfo {author} {\bibfnamefont {A.}~\bibnamefont {Sorensen}},\ and\ \bibinfo {author} {\bibfnamefont {M.}~\bibnamefont {Stephanov}},\ }\bibfield  {title} {\bibinfo {title} {{The QCD phase diagram and Beam Energy Scan physics: A theory overview}},\ }\href {https://doi.org/10.1142/9789811294679_0007} {\bibfield  {journal} {\bibinfo  {journal} {Int. J. Mod. Phys. E}\ }\textbf {\bibinfo {volume} {33}},\ \bibinfo {pages} {2430008} (\bibinfo {year} {2024})},\ \Eprint {https://arxiv.org/abs/2402.10183} {arXiv:2402.10183 [nucl-th]} \BibitemShut {NoStop}%
\bibitem [{\citenamefont {Savchuk}(2025{\natexlab{a}})}]{Savchuk:2024ykb}%
  \BibitemOpen
  \bibfield  {author} {\bibinfo {author} {\bibfnamefont {O.}~\bibnamefont {Savchuk}},\ }\bibfield  {title} {\bibinfo {title} {{Net-proton fluctuations influenced by baryon stopping and quark deconfinement}},\ }\href {https://doi.org/10.1103/PhysRevC.111.024913} {\bibfield  {journal} {\bibinfo  {journal} {Phys. Rev. C}\ }\textbf {\bibinfo {volume} {111}},\ \bibinfo {pages} {024913} (\bibinfo {year} {2025}{\natexlab{a}})},\ \Eprint {https://arxiv.org/abs/2407.17670} {arXiv:2407.17670 [hep-ph]} \BibitemShut {NoStop}%
\bibitem [{\citenamefont {Sorensen}\ \emph {et~al.}(2024)\citenamefont {Sorensen} \emph {et~al.}}]{Sorensen:2023zkk}%
  \BibitemOpen
  \bibfield  {author} {\bibinfo {author} {\bibfnamefont {A.}~\bibnamefont {Sorensen}} \emph {et~al.},\ }\bibfield  {title} {\bibinfo {title} {{Dense nuclear matter equation of state from heavy-ion collisions}},\ }\href {https://doi.org/10.1016/j.ppnp.2023.104080} {\bibfield  {journal} {\bibinfo  {journal} {Prog. Part. Nucl. Phys.}\ }\textbf {\bibinfo {volume} {134}},\ \bibinfo {pages} {104080} (\bibinfo {year} {2024})},\ \Eprint {https://arxiv.org/abs/2301.13253} {arXiv:2301.13253 [nucl-th]} \BibitemShut {NoStop}%
\bibitem [{\citenamefont {Brown}\ \emph {et~al.}(2025)\citenamefont {Brown} \emph {et~al.}}]{Brown:2024rml}%
  \BibitemOpen
  \bibfield  {author} {\bibinfo {author} {\bibfnamefont {B.~A.}\ \bibnamefont {Brown}} \emph {et~al.},\ }\bibfield  {title} {\bibinfo {title} {{Motivations for early high-profile FRIB experiments}},\ }\href {https://doi.org/10.1088/1361-6471/adb449} {\bibfield  {journal} {\bibinfo  {journal} {J. Phys. G}\ }\textbf {\bibinfo {volume} {52}},\ \bibinfo {pages} {050501} (\bibinfo {year} {2025})},\ \Eprint {https://arxiv.org/abs/2410.06144} {arXiv:2410.06144 [nucl-th]} \BibitemShut {NoStop}%
\bibitem [{\citenamefont {Lynn}\ \emph {et~al.}(2019)\citenamefont {Lynn}, \citenamefont {Tews}, \citenamefont {Gandolfi},\ and\ \citenamefont {Lovato}}]{Lynn:2019rdt}%
  \BibitemOpen
  \bibfield  {author} {\bibinfo {author} {\bibfnamefont {J.~E.}\ \bibnamefont {Lynn}}, \bibinfo {author} {\bibfnamefont {I.}~\bibnamefont {Tews}}, \bibinfo {author} {\bibfnamefont {S.}~\bibnamefont {Gandolfi}},\ and\ \bibinfo {author} {\bibfnamefont {A.}~\bibnamefont {Lovato}},\ }\bibfield  {title} {\bibinfo {title} {{Quantum Monte Carlo Methods in Nuclear Physics: Recent Advances}},\ }\href {https://doi.org/10.1146/annurev-nucl-101918-023600} {\bibfield  {journal} {\bibinfo  {journal} {Ann. Rev. Nucl. Part. Sci.}\ }\textbf {\bibinfo {volume} {69}},\ \bibinfo {pages} {279} (\bibinfo {year} {2019})},\ \Eprint {https://arxiv.org/abs/1901.04868} {arXiv:1901.04868 [nucl-th]} \BibitemShut {NoStop}%
\bibitem [{\citenamefont {Drischler}\ \emph {et~al.}(2021)\citenamefont {Drischler}, \citenamefont {Holt},\ and\ \citenamefont {Wellenhofer}}]{Drischler:2021kxf}%
  \BibitemOpen
  \bibfield  {author} {\bibinfo {author} {\bibfnamefont {C.}~\bibnamefont {Drischler}}, \bibinfo {author} {\bibfnamefont {J.~W.}\ \bibnamefont {Holt}},\ and\ \bibinfo {author} {\bibfnamefont {C.}~\bibnamefont {Wellenhofer}},\ }\bibfield  {title} {\bibinfo {title} {{Chiral Effective Field Theory and the High-Density Nuclear Equation of State}},\ }\href {https://doi.org/10.1146/annurev-nucl-102419-041903} {\bibfield  {journal} {\bibinfo  {journal} {Ann. Rev. Nucl. Part. Sci.}\ }\textbf {\bibinfo {volume} {71}},\ \bibinfo {pages} {403} (\bibinfo {year} {2021})},\ \Eprint {https://arxiv.org/abs/2101.01709} {arXiv:2101.01709 [nucl-th]} \BibitemShut {NoStop}%
\bibitem [{\citenamefont {{\"O}zel}\ and\ \citenamefont {Freire}(2016)}]{Ozel:2016oaf}%
  \BibitemOpen
  \bibfield  {author} {\bibinfo {author} {\bibfnamefont {F.}~\bibnamefont {{\"O}zel}}\ and\ \bibinfo {author} {\bibfnamefont {P.}~\bibnamefont {Freire}},\ }\bibfield  {title} {\bibinfo {title} {{Masses, Radii, and the Equation of State of Neutron Stars}},\ }\href {https://doi.org/10.1146/annurev-astro-081915-023322} {\bibfield  {journal} {\bibinfo  {journal} {Ann. Rev. Astron. Astrophys.}\ }\textbf {\bibinfo {volume} {54}},\ \bibinfo {pages} {401} (\bibinfo {year} {2016})},\ \Eprint {https://arxiv.org/abs/1603.02698} {arXiv:1603.02698 [astro-ph.HE]} \BibitemShut {NoStop}%
\bibitem [{\citenamefont {Lattimer}(2021)}]{Lattimer:2021emm}%
  \BibitemOpen
  \bibfield  {author} {\bibinfo {author} {\bibfnamefont {J.~M.}\ \bibnamefont {Lattimer}},\ }\bibfield  {title} {\bibinfo {title} {{Neutron Stars and the Nuclear Matter Equation of State}},\ }\href {https://doi.org/10.1146/annurev-nucl-102419-124827} {\bibfield  {journal} {\bibinfo  {journal} {Ann. Rev. Nucl. Part. Sci.}\ }\textbf {\bibinfo {volume} {71}},\ \bibinfo {pages} {433} (\bibinfo {year} {2021})}\BibitemShut {NoStop}%
\bibitem [{\citenamefont {Essick}\ \emph {et~al.}(2021)\citenamefont {Essick}, \citenamefont {Tews}, \citenamefont {Landry},\ and\ \citenamefont {Schwenk}}]{Essick:2021kjb}%
  \BibitemOpen
  \bibfield  {author} {\bibinfo {author} {\bibfnamefont {R.}~\bibnamefont {Essick}}, \bibinfo {author} {\bibfnamefont {I.}~\bibnamefont {Tews}}, \bibinfo {author} {\bibfnamefont {P.}~\bibnamefont {Landry}},\ and\ \bibinfo {author} {\bibfnamefont {A.}~\bibnamefont {Schwenk}},\ }\bibfield  {title} {\bibinfo {title} {{Astrophysical Constraints on the Symmetry Energy and the Neutron Skin of Pb208 with Minimal Modeling Assumptions}},\ }\href {https://doi.org/10.1103/PhysRevLett.127.192701} {\bibfield  {journal} {\bibinfo  {journal} {Phys. Rev. Lett.}\ }\textbf {\bibinfo {volume} {127}},\ \bibinfo {pages} {192701} (\bibinfo {year} {2021})},\ \Eprint {https://arxiv.org/abs/2102.10074} {arXiv:2102.10074 [nucl-th]} \BibitemShut {NoStop}%
\bibitem [{\citenamefont {Dietrich}\ \emph {et~al.}(2020)\citenamefont {Dietrich}, \citenamefont {Coughlin}, \citenamefont {Pang}, \citenamefont {Bulla}, \citenamefont {Heinzel}, \citenamefont {Issa}, \citenamefont {Tews},\ and\ \citenamefont {Antier}}]{Dietrich:2020efo}%
  \BibitemOpen
  \bibfield  {author} {\bibinfo {author} {\bibfnamefont {T.}~\bibnamefont {Dietrich}}, \bibinfo {author} {\bibfnamefont {M.~W.}\ \bibnamefont {Coughlin}}, \bibinfo {author} {\bibfnamefont {P.~T.~H.}\ \bibnamefont {Pang}}, \bibinfo {author} {\bibfnamefont {M.}~\bibnamefont {Bulla}}, \bibinfo {author} {\bibfnamefont {J.}~\bibnamefont {Heinzel}}, \bibinfo {author} {\bibfnamefont {L.}~\bibnamefont {Issa}}, \bibinfo {author} {\bibfnamefont {I.}~\bibnamefont {Tews}},\ and\ \bibinfo {author} {\bibfnamefont {S.}~\bibnamefont {Antier}},\ }\bibfield  {title} {\bibinfo {title} {{Multimessenger constraints on the neutron-star equation of state and the Hubble constant}},\ }\href {https://doi.org/10.1126/science.abb4317} {\bibfield  {journal} {\bibinfo  {journal} {Science}\ }\textbf {\bibinfo {volume} {370}},\ \bibinfo {pages} {1450} (\bibinfo {year} {2020})},\ \Eprint {https://arxiv.org/abs/2002.11355} {arXiv:2002.11355 [astro-ph.HE]} \BibitemShut {NoStop}%
\bibitem [{\citenamefont {Shuryak}(1978)}]{Shuryak:1978ij}%
  \BibitemOpen
  \bibfield  {author} {\bibinfo {author} {\bibfnamefont {E.~V.}\ \bibnamefont {Shuryak}},\ }\bibfield  {title} {\bibinfo {title} {{Quark-Gluon Plasma and Hadronic Production of Leptons, Photons and Psions}},\ }\href {https://doi.org/10.1016/0370-2693(78)90370-2} {\bibfield  {journal} {\bibinfo  {journal} {Phys. Lett. B}\ }\textbf {\bibinfo {volume} {78}},\ \bibinfo {pages} {150} (\bibinfo {year} {1978})}\BibitemShut {NoStop}%
\bibitem [{\citenamefont {McLerran}\ and\ \citenamefont {Toimela}(1985)}]{McLerran:1984ay}%
  \BibitemOpen
  \bibfield  {author} {\bibinfo {author} {\bibfnamefont {L.~D.}\ \bibnamefont {McLerran}}\ and\ \bibinfo {author} {\bibfnamefont {T.}~\bibnamefont {Toimela}},\ }\bibfield  {title} {\bibinfo {title} {{Photon and Dilepton Emission from the Quark - Gluon Plasma: Some General Considerations}},\ }\href {https://doi.org/10.1103/PhysRevD.31.545} {\bibfield  {journal} {\bibinfo  {journal} {Phys. Rev. D}\ }\textbf {\bibinfo {volume} {31}},\ \bibinfo {pages} {545} (\bibinfo {year} {1985})}\BibitemShut {NoStop}%
\bibitem [{\citenamefont {Rapp}\ and\ \citenamefont {Wambach}(2000)}]{Rapp:1999ej}%
  \BibitemOpen
  \bibfield  {author} {\bibinfo {author} {\bibfnamefont {R.}~\bibnamefont {Rapp}}\ and\ \bibinfo {author} {\bibfnamefont {J.}~\bibnamefont {Wambach}},\ }\bibfield  {title} {\bibinfo {title} {{Chiral symmetry restoration and dileptons in relativistic heavy ion collisions}},\ }\href {https://doi.org/10.1007/0-306-47101-9_1} {\bibfield  {journal} {\bibinfo  {journal} {Adv. Nucl. Phys.}\ }\textbf {\bibinfo {volume} {25}},\ \bibinfo {pages} {1} (\bibinfo {year} {2000})},\ \Eprint {https://arxiv.org/abs/hep-ph/9909229} {arXiv:hep-ph/9909229} \BibitemShut {NoStop}%
\bibitem [{\citenamefont {Danielewicz}\ and\ \citenamefont {Odyniec}(1985)}]{Danielewicz:1985hn}%
  \BibitemOpen
  \bibfield  {author} {\bibinfo {author} {\bibfnamefont {P.}~\bibnamefont {Danielewicz}}\ and\ \bibinfo {author} {\bibfnamefont {G.}~\bibnamefont {Odyniec}},\ }\bibfield  {title} {\bibinfo {title} {{Transverse Momentum Analysis of Collective Motion in Relativistic Nuclear Collisions}},\ }\href {https://doi.org/10.1016/0370-2693(85)91535-7} {\bibfield  {journal} {\bibinfo  {journal} {Phys. Lett. B}\ }\textbf {\bibinfo {volume} {157}},\ \bibinfo {pages} {146} (\bibinfo {year} {1985})},\ \Eprint {https://arxiv.org/abs/2109.05308} {arXiv:2109.05308 [nucl-th]} \BibitemShut {NoStop}%
\bibitem [{\citenamefont {Rischke}\ \emph {et~al.}(1995)\citenamefont {Rischke}, \citenamefont {P{\"u}rs{\"u}n}, \citenamefont {Maruhn}, \citenamefont {Stoecker},\ and\ \citenamefont {Greiner}}]{Rischke:1995pe}%
  \BibitemOpen
  \bibfield  {author} {\bibinfo {author} {\bibfnamefont {D.~H.}\ \bibnamefont {Rischke}}, \bibinfo {author} {\bibfnamefont {Y.}~\bibnamefont {P{\"u}rs{\"u}n}}, \bibinfo {author} {\bibfnamefont {J.~A.}\ \bibnamefont {Maruhn}}, \bibinfo {author} {\bibfnamefont {H.}~\bibnamefont {Stoecker}},\ and\ \bibinfo {author} {\bibfnamefont {W.}~\bibnamefont {Greiner}},\ }\bibfield  {title} {\bibinfo {title} {{The Phase transition to the quark - gluon plasma and its effects on hydrodynamic flow}},\ }\href {https://doi.org/10.1007/BF03053749} {\bibfield  {journal} {\bibinfo  {journal} {Acta Phys. Hung. A}\ }\textbf {\bibinfo {volume} {1}},\ \bibinfo {pages} {309} (\bibinfo {year} {1995})},\ \Eprint {https://arxiv.org/abs/nucl-th/9505014} {arXiv:nucl-th/9505014} \BibitemShut {NoStop}%
\bibitem [{\citenamefont {Stoecker}(2005)}]{Stoecker:2004qu}%
  \BibitemOpen
  \bibfield  {author} {\bibinfo {author} {\bibfnamefont {H.}~\bibnamefont {Stoecker}},\ }\bibfield  {title} {\bibinfo {title} {{Collective flow signals the quark gluon plasma}},\ }\href {https://doi.org/10.1016/j.nuclphysa.2004.12.074} {\bibfield  {journal} {\bibinfo  {journal} {Nucl. Phys. A}\ }\textbf {\bibinfo {volume} {750}},\ \bibinfo {pages} {121} (\bibinfo {year} {2005})},\ \Eprint {https://arxiv.org/abs/nucl-th/0406018} {arXiv:nucl-th/0406018} \BibitemShut {NoStop}%
\bibitem [{\citenamefont {Brachmann}\ \emph {et~al.}(2000{\natexlab{a}})\citenamefont {Brachmann}, \citenamefont {Soff}, \citenamefont {Dumitru}, \citenamefont {Stoecker}, \citenamefont {Maruhn}, \citenamefont {Greiner}, \citenamefont {Bravina},\ and\ \citenamefont {Rischke}}]{Brachmann:1999xt}%
  \BibitemOpen
  \bibfield  {author} {\bibinfo {author} {\bibfnamefont {J.}~\bibnamefont {Brachmann}}, \bibinfo {author} {\bibfnamefont {S.}~\bibnamefont {Soff}}, \bibinfo {author} {\bibfnamefont {A.}~\bibnamefont {Dumitru}}, \bibinfo {author} {\bibfnamefont {H.}~\bibnamefont {Stoecker}}, \bibinfo {author} {\bibfnamefont {J.~A.}\ \bibnamefont {Maruhn}}, \bibinfo {author} {\bibfnamefont {W.}~\bibnamefont {Greiner}}, \bibinfo {author} {\bibfnamefont {L.~V.}\ \bibnamefont {Bravina}},\ and\ \bibinfo {author} {\bibfnamefont {D.~H.}\ \bibnamefont {Rischke}},\ }\bibfield  {title} {\bibinfo {title} {{Antiflow of nucleons at the softest point of the EoS}},\ }\href {https://doi.org/10.1103/PhysRevC.61.024909} {\bibfield  {journal} {\bibinfo  {journal} {Phys. Rev. C}\ }\textbf {\bibinfo {volume} {61}},\ \bibinfo {pages} {024909} (\bibinfo {year} {2000}{\natexlab{a}})},\ \Eprint {https://arxiv.org/abs/nucl-th/9908010} {arXiv:nucl-th/9908010} \BibitemShut {NoStop}%
\bibitem [{\citenamefont {Brachmann}\ \emph {et~al.}(2000{\natexlab{b}})\citenamefont {Brachmann}, \citenamefont {Dumitru}, \citenamefont {Stoecker},\ and\ \citenamefont {Greiner}}]{Brachmann:1999mp}%
  \BibitemOpen
  \bibfield  {author} {\bibinfo {author} {\bibfnamefont {J.}~\bibnamefont {Brachmann}}, \bibinfo {author} {\bibfnamefont {A.}~\bibnamefont {Dumitru}}, \bibinfo {author} {\bibfnamefont {H.}~\bibnamefont {Stoecker}},\ and\ \bibinfo {author} {\bibfnamefont {W.}~\bibnamefont {Greiner}},\ }\bibfield  {title} {\bibinfo {title} {{The Directed flow maximum near c(s) = 0}},\ }\href {https://doi.org/10.1007/s100500070077} {\bibfield  {journal} {\bibinfo  {journal} {Eur. Phys. J. A}\ }\textbf {\bibinfo {volume} {8}},\ \bibinfo {pages} {549} (\bibinfo {year} {2000}{\natexlab{b}})},\ \Eprint {https://arxiv.org/abs/nucl-th/9912014} {arXiv:nucl-th/9912014} \BibitemShut {NoStop}%
\bibitem [{\citenamefont {Oliinychenko}\ \emph {et~al.}(2023)\citenamefont {Oliinychenko}, \citenamefont {Sorensen}, \citenamefont {Koch},\ and\ \citenamefont {McLerran}}]{Oliinychenko:2022uvy}%
  \BibitemOpen
  \bibfield  {author} {\bibinfo {author} {\bibfnamefont {D.}~\bibnamefont {Oliinychenko}}, \bibinfo {author} {\bibfnamefont {A.}~\bibnamefont {Sorensen}}, \bibinfo {author} {\bibfnamefont {V.}~\bibnamefont {Koch}},\ and\ \bibinfo {author} {\bibfnamefont {L.}~\bibnamefont {McLerran}},\ }\bibfield  {title} {\bibinfo {title} {{Sensitivity of Au+Au collisions to the symmetric nuclear matter equation~ of state at 2{\textendash}5 nuclear saturation densities}},\ }\href {https://doi.org/10.1103/PhysRevC.108.034908} {\bibfield  {journal} {\bibinfo  {journal} {Phys. Rev. C}\ }\textbf {\bibinfo {volume} {108}},\ \bibinfo {pages} {034908} (\bibinfo {year} {2023})},\ \Eprint {https://arxiv.org/abs/2208.11996} {arXiv:2208.11996 [nucl-th]} \BibitemShut {NoStop}%
\bibitem [{\citenamefont {Adamczyk}\ \emph {et~al.}(2018)\citenamefont {Adamczyk} \emph {et~al.}}]{STAR:2017ykf}%
  \BibitemOpen
  \bibfield  {author} {\bibinfo {author} {\bibfnamefont {L.}~\bibnamefont {Adamczyk}} \emph {et~al.} (\bibinfo {collaboration} {STAR}),\ }\bibfield  {title} {\bibinfo {title} {{Azimuthal anisotropy in Cu$+$Au collisions at $\sqrt{s_{_{NN}}}$ = 200 GeV}},\ }\href {https://doi.org/10.1103/PhysRevC.98.014915} {\bibfield  {journal} {\bibinfo  {journal} {Phys. Rev. C}\ }\textbf {\bibinfo {volume} {98}},\ \bibinfo {pages} {014915} (\bibinfo {year} {2018})},\ \Eprint {https://arxiv.org/abs/1712.01332} {arXiv:1712.01332 [nucl-ex]} \BibitemShut {NoStop}%
\bibitem [{\citenamefont {Lednicky}\ \emph {et~al.}(1996)\citenamefont {Lednicky}, \citenamefont {Lyuboshits}, \citenamefont {Erazmus},\ and\ \citenamefont {Nouais}}]{Lednicky:1995vk}%
  \BibitemOpen
  \bibfield  {author} {\bibinfo {author} {\bibfnamefont {R.}~\bibnamefont {Lednicky}}, \bibinfo {author} {\bibfnamefont {V.~L.}\ \bibnamefont {Lyuboshits}}, \bibinfo {author} {\bibfnamefont {B.}~\bibnamefont {Erazmus}},\ and\ \bibinfo {author} {\bibfnamefont {D.}~\bibnamefont {Nouais}},\ }\bibfield  {title} {\bibinfo {title} {{How to measure which sort of particles was emitted earlier and which later}},\ }\href {https://doi.org/10.1016/0370-2693(96)00124-4} {\bibfield  {journal} {\bibinfo  {journal} {Phys. Lett. B}\ }\textbf {\bibinfo {volume} {373}},\ \bibinfo {pages} {30} (\bibinfo {year} {1996})}\BibitemShut {NoStop}%
\bibitem [{\citenamefont {Lednicky}(2009)}]{Lednicky:2005tb}%
  \BibitemOpen
  \bibfield  {author} {\bibinfo {author} {\bibfnamefont {R.}~\bibnamefont {Lednicky}},\ }\bibfield  {title} {\bibinfo {title} {{Finite-size effects on two-particle production in continuous and discrete spectrum}},\ }\href {https://doi.org/10.1134/S1063779609030034} {\bibfield  {journal} {\bibinfo  {journal} {Phys. Part. Nucl.}\ }\textbf {\bibinfo {volume} {40}},\ \bibinfo {pages} {307} (\bibinfo {year} {2009})},\ \Eprint {https://arxiv.org/abs/nucl-th/0501065} {arXiv:nucl-th/0501065} \BibitemShut {NoStop}%
\bibitem [{\citenamefont {Pratt}(1984)}]{Pratt:1984su}%
  \BibitemOpen
  \bibfield  {author} {\bibinfo {author} {\bibfnamefont {S.}~\bibnamefont {Pratt}},\ }\bibfield  {title} {\bibinfo {title} {{Pion Interferometry for Exploding Sources}},\ }\href {https://doi.org/10.1103/PhysRevLett.53.1219} {\bibfield  {journal} {\bibinfo  {journal} {Phys. Rev. Lett.}\ }\textbf {\bibinfo {volume} {53}},\ \bibinfo {pages} {1219} (\bibinfo {year} {1984})}\BibitemShut {NoStop}%
\bibitem [{\citenamefont {Wiedemann}(1998)}]{Wiedemann:1997cr}%
  \BibitemOpen
  \bibfield  {author} {\bibinfo {author} {\bibfnamefont {U.~A.}\ \bibnamefont {Wiedemann}},\ }\bibfield  {title} {\bibinfo {title} {{Two particle interferometry for noncentral heavy ion collisions}},\ }\href {https://doi.org/10.1103/PhysRevC.57.266} {\bibfield  {journal} {\bibinfo  {journal} {Phys. Rev. C}\ }\textbf {\bibinfo {volume} {57}},\ \bibinfo {pages} {266} (\bibinfo {year} {1998})},\ \Eprint {https://arxiv.org/abs/nucl-th/9707046} {arXiv:nucl-th/9707046} \BibitemShut {NoStop}%
\bibitem [{\citenamefont {Khyzhniak}\ and\ \citenamefont {Lisa}(2025)}]{Khyzhniak:2024chj}%
  \BibitemOpen
  \bibfield  {author} {\bibinfo {author} {\bibfnamefont {Y.}~\bibnamefont {Khyzhniak}}\ and\ \bibinfo {author} {\bibfnamefont {M.~A.}\ \bibnamefont {Lisa}},\ }\bibfield  {title} {\bibinfo {title} {{Pair momentum dependence of a tilted source in heavy-ion collisions}},\ }\href {https://doi.org/10.1103/PhysRevC.111.024902} {\bibfield  {journal} {\bibinfo  {journal} {Phys. Rev. C}\ }\textbf {\bibinfo {volume} {111}},\ \bibinfo {pages} {024902} (\bibinfo {year} {2025})},\ \Eprint {https://arxiv.org/abs/2410.15134} {arXiv:2410.15134 [nucl-th]} \BibitemShut {NoStop}%
\bibitem [{\citenamefont {Becattini}\ and\ \citenamefont {Lisa}(2020)}]{Becattini:2020ngo}%
  \BibitemOpen
  \bibfield  {author} {\bibinfo {author} {\bibfnamefont {F.}~\bibnamefont {Becattini}}\ and\ \bibinfo {author} {\bibfnamefont {M.~A.}\ \bibnamefont {Lisa}},\ }\bibfield  {title} {\bibinfo {title} {{Polarization and Vorticity in the Quark{\textendash}Gluon Plasma}},\ }\href {https://doi.org/10.1146/annurev-nucl-021920-095245} {\bibfield  {journal} {\bibinfo  {journal} {Ann. Rev. Nucl. Part. Sci.}\ }\textbf {\bibinfo {volume} {70}},\ \bibinfo {pages} {395} (\bibinfo {year} {2020})},\ \Eprint {https://arxiv.org/abs/2003.03640} {arXiv:2003.03640 [nucl-ex]} \BibitemShut {NoStop}%
\bibitem [{\citenamefont {Florkowski}\ \emph {et~al.}(2018)\citenamefont {Florkowski}, \citenamefont {Kumar},\ and\ \citenamefont {Ryblewski}}]{Florkowski:2018ahw}%
  \BibitemOpen
  \bibfield  {author} {\bibinfo {author} {\bibfnamefont {W.}~\bibnamefont {Florkowski}}, \bibinfo {author} {\bibfnamefont {A.}~\bibnamefont {Kumar}},\ and\ \bibinfo {author} {\bibfnamefont {R.}~\bibnamefont {Ryblewski}},\ }\bibfield  {title} {\bibinfo {title} {{Thermodynamic versus kinetic approach to polarization-vorticity coupling}},\ }\href {https://doi.org/10.1103/PhysRevC.98.044906} {\bibfield  {journal} {\bibinfo  {journal} {Phys. Rev. C}\ }\textbf {\bibinfo {volume} {98}},\ \bibinfo {pages} {044906} (\bibinfo {year} {2018})},\ \Eprint {https://arxiv.org/abs/1806.02616} {arXiv:1806.02616 [hep-ph]} \BibitemShut {NoStop}%
\bibitem [{\citenamefont {Palermo}\ \emph {et~al.}(2024)\citenamefont {Palermo}, \citenamefont {Grossi}, \citenamefont {Karpenko},\ and\ \citenamefont {Becattini}}]{Palermo:2024tza}%
  \BibitemOpen
  \bibfield  {author} {\bibinfo {author} {\bibfnamefont {A.}~\bibnamefont {Palermo}}, \bibinfo {author} {\bibfnamefont {E.}~\bibnamefont {Grossi}}, \bibinfo {author} {\bibfnamefont {I.}~\bibnamefont {Karpenko}},\ and\ \bibinfo {author} {\bibfnamefont {F.}~\bibnamefont {Becattini}},\ }\bibfield  {title} {\bibinfo {title} {{$\Lambda $ polarization in very high energy heavy ion collisions as a probe of the quark{\textendash}gluon plasma formation and properties}},\ }\href {https://doi.org/10.1140/epjc/s10052-024-13229-z} {\bibfield  {journal} {\bibinfo  {journal} {Eur. Phys. J. C}\ }\textbf {\bibinfo {volume} {84}},\ \bibinfo {pages} {920} (\bibinfo {year} {2024})},\ \Eprint {https://arxiv.org/abs/2404.14295} {arXiv:2404.14295 [nucl-th]} \BibitemShut {NoStop}%
\bibitem [{\citenamefont {Adamczyk}\ \emph {et~al.}(2017{\natexlab{b}})\citenamefont {Adamczyk} \emph {et~al.}}]{STAR:2017ckg}%
  \BibitemOpen
  \bibfield  {author} {\bibinfo {author} {\bibfnamefont {L.}~\bibnamefont {Adamczyk}} \emph {et~al.} (\bibinfo {collaboration} {STAR}),\ }\bibfield  {title} {\bibinfo {title} {{Global $\Lambda$ hyperon polarization in nuclear collisions: evidence for the most vortical fluid}},\ }\href {https://doi.org/10.1038/nature23004} {\bibfield  {journal} {\bibinfo  {journal} {Nature}\ }\textbf {\bibinfo {volume} {548}},\ \bibinfo {pages} {62} (\bibinfo {year} {2017}{\natexlab{b}})},\ \Eprint {https://arxiv.org/abs/1701.06657} {arXiv:1701.06657 [nucl-ex]} \BibitemShut {NoStop}%
\bibitem [{\citenamefont {Abou~Yassine}\ \emph {et~al.}(2022)\citenamefont {Abou~Yassine} \emph {et~al.}}]{HADES:2022enx}%
  \BibitemOpen
  \bibfield  {author} {\bibinfo {author} {\bibfnamefont {R.}~\bibnamefont {Abou~Yassine}} \emph {et~al.} (\bibinfo {collaboration} {HADES}),\ }\bibfield  {title} {\bibinfo {title} {{Measurement of global polarization of {\ensuremath{\Lambda}} hyperons in few-GeV heavy-ion collisions}},\ }\href {https://doi.org/10.1016/j.physletb.2022.137506} {\bibfield  {journal} {\bibinfo  {journal} {Phys. Lett. B}\ }\textbf {\bibinfo {volume} {835}},\ \bibinfo {pages} {137506} (\bibinfo {year} {2022})},\ \Eprint {https://arxiv.org/abs/2207.05160} {arXiv:2207.05160 [nucl-ex]} \BibitemShut {NoStop}%
\bibitem [{\citenamefont {Giacalone}\ and\ \citenamefont {Speranza}(2025)}]{Giacalone:2025bgm}%
  \BibitemOpen
  \bibfield  {author} {\bibinfo {author} {\bibfnamefont {G.}~\bibnamefont {Giacalone}}\ and\ \bibinfo {author} {\bibfnamefont {E.}~\bibnamefont {Speranza}},\ }\bibfield  {title} {\bibinfo {title} {Initial-state-driven spin correlations in high-energy nuclear collisions},\ }\href@noop {} {\bibfield  {journal} {\bibinfo  {journal} {arXiv preprint arXiv:2502.13102}\ } (\bibinfo {year} {2025})}\BibitemShut {NoStop}%
\bibitem [{\citenamefont {Becattini}\ \emph {et~al.}(2013)\citenamefont {Becattini}, \citenamefont {Chandra}, \citenamefont {Del~Zanna},\ and\ \citenamefont {Grossi}}]{Becattini:2013fla}%
  \BibitemOpen
  \bibfield  {author} {\bibinfo {author} {\bibfnamefont {F.}~\bibnamefont {Becattini}}, \bibinfo {author} {\bibfnamefont {V.}~\bibnamefont {Chandra}}, \bibinfo {author} {\bibfnamefont {L.}~\bibnamefont {Del~Zanna}},\ and\ \bibinfo {author} {\bibfnamefont {E.}~\bibnamefont {Grossi}},\ }\bibfield  {title} {\bibinfo {title} {{Relativistic distribution function for particles with spin at local thermodynamical equilibrium}},\ }\href {https://doi.org/10.1016/j.aop.2013.07.004} {\bibfield  {journal} {\bibinfo  {journal} {Annals Phys.}\ }\textbf {\bibinfo {volume} {338}},\ \bibinfo {pages} {32} (\bibinfo {year} {2013})},\ \Eprint {https://arxiv.org/abs/1303.3431} {arXiv:1303.3431 [nucl-th]} \BibitemShut {NoStop}%
\bibitem [{\citenamefont {Koonin}(1977)}]{Koonin:1977fh}%
  \BibitemOpen
  \bibfield  {author} {\bibinfo {author} {\bibfnamefont {S.~E.}\ \bibnamefont {Koonin}},\ }\bibfield  {title} {\bibinfo {title} {{Proton Pictures of High-Energy Nuclear Collisions}},\ }\href {https://doi.org/10.1016/0370-2693(77)90340-9} {\bibfield  {journal} {\bibinfo  {journal} {Phys. Lett. B}\ }\textbf {\bibinfo {volume} {70}},\ \bibinfo {pages} {43} (\bibinfo {year} {1977})}\BibitemShut {NoStop}%
\bibitem [{\citenamefont {Pratt}(1986)}]{Pratt:1986cc}%
  \BibitemOpen
  \bibfield  {author} {\bibinfo {author} {\bibfnamefont {S.}~\bibnamefont {Pratt}},\ }\bibfield  {title} {\bibinfo {title} {{Pion Interferometry of Quark-Gluon Plasma}},\ }\href {https://doi.org/10.1103/PhysRevD.33.1314} {\bibfield  {journal} {\bibinfo  {journal} {Phys. Rev. D}\ }\textbf {\bibinfo {volume} {33}},\ \bibinfo {pages} {1314} (\bibinfo {year} {1986})}\BibitemShut {NoStop}%
\bibitem [{\citenamefont {Lisa}\ \emph {et~al.}(2005)\citenamefont {Lisa}, \citenamefont {Pratt}, \citenamefont {Soltz},\ and\ \citenamefont {Wiedemann}}]{Lisa:2005dd}%
  \BibitemOpen
  \bibfield  {author} {\bibinfo {author} {\bibfnamefont {M.~A.}\ \bibnamefont {Lisa}}, \bibinfo {author} {\bibfnamefont {S.}~\bibnamefont {Pratt}}, \bibinfo {author} {\bibfnamefont {R.}~\bibnamefont {Soltz}},\ and\ \bibinfo {author} {\bibfnamefont {U.}~\bibnamefont {Wiedemann}},\ }\bibfield  {title} {\bibinfo {title} {{Femtoscopy in relativistic heavy ion collisions}},\ }\href {https://doi.org/10.1146/annurev.nucl.55.090704.151533} {\bibfield  {journal} {\bibinfo  {journal} {Ann. Rev. Nucl. Part. Sci.}\ }\textbf {\bibinfo {volume} {55}},\ \bibinfo {pages} {357} (\bibinfo {year} {2005})},\ \Eprint {https://arxiv.org/abs/nucl-ex/0505014} {arXiv:nucl-ex/0505014} \BibitemShut {NoStop}%
\bibitem [{\citenamefont {Pratt}(1997)}]{Pratt:1997pw}%
  \BibitemOpen
  \bibfield  {author} {\bibinfo {author} {\bibfnamefont {S.}~\bibnamefont {Pratt}},\ }\bibfield  {title} {\bibinfo {title} {{Validity of the smoothness assumption for calculating two-boson correlations in high-energy collisions}},\ }\href {https://doi.org/10.1103/PhysRevC.56.1095} {\bibfield  {journal} {\bibinfo  {journal} {Phys. Rev. C}\ }\textbf {\bibinfo {volume} {56}},\ \bibinfo {pages} {1095} (\bibinfo {year} {1997})}\BibitemShut {NoStop}%
\bibitem [{\citenamefont {Rz{\k{e}}sa}\ \emph {et~al.}(2025)\citenamefont {Rz{\k{e}}sa}, \citenamefont {Stefaniak},\ and\ \citenamefont {Pratt}}]{Rzesa:2024oqp}%
  \BibitemOpen
  \bibfield  {author} {\bibinfo {author} {\bibfnamefont {W.}~\bibnamefont {Rz{\k{e}}sa}}, \bibinfo {author} {\bibfnamefont {M.}~\bibnamefont {Stefaniak}},\ and\ \bibinfo {author} {\bibfnamefont {S.}~\bibnamefont {Pratt}},\ }\bibfield  {title} {\bibinfo {title} {{Theoretical description of proton-deuteron interactions using exact two-body dynamics of the femtoscopic correlation method}},\ }\href {https://doi.org/10.1103/PhysRevC.111.034903} {\bibfield  {journal} {\bibinfo  {journal} {Phys. Rev. C}\ }\textbf {\bibinfo {volume} {111}},\ \bibinfo {pages} {034903} (\bibinfo {year} {2025})},\ \Eprint {https://arxiv.org/abs/2410.13983} {arXiv:2410.13983 [nucl-th]} \BibitemShut {NoStop}%
\bibitem [{\citenamefont {Adams}\ \emph {et~al.}(2006)\citenamefont {Adams} \emph {et~al.}}]{STAR:2005rpl}%
  \BibitemOpen
  \bibfield  {author} {\bibinfo {author} {\bibfnamefont {J.}~\bibnamefont {Adams}} \emph {et~al.} (\bibinfo {collaboration} {STAR}),\ }\bibfield  {title} {\bibinfo {title} {{Proton - lambda correlations in central Au+Au collisions at S(NN)**(1/2) = 200-GeV}},\ }\href {https://doi.org/10.1103/PhysRevC.74.064906} {\bibfield  {journal} {\bibinfo  {journal} {Phys. Rev. C}\ }\textbf {\bibinfo {volume} {74}},\ \bibinfo {pages} {064906} (\bibinfo {year} {2006})},\ \Eprint {https://arxiv.org/abs/nucl-ex/0511003} {arXiv:nucl-ex/0511003} \BibitemShut {NoStop}%
\bibitem [{\citenamefont {Stefaniak}(2024{\natexlab{a}})}]{Stefaniak:2024fkf}%
  \BibitemOpen
  \bibfield  {author} {\bibinfo {author} {\bibfnamefont {M.}~\bibnamefont {Stefaniak}},\ }\bibfield  {title} {\bibinfo {title} {Proton-proton and proton-cluster femtoscopy at the hades experiment},\ }\href@noop {} {\bibfield  {journal} {\bibinfo  {journal} {arXiv preprint arXiv:2402.09280}\ } (\bibinfo {year} {2024}{\natexlab{a}})}\BibitemShut {NoStop}%
\bibitem [{\citenamefont {Stefaniak}(2024{\natexlab{b}})}]{Stefaniak:2024eux}%
  \BibitemOpen
  \bibfield  {author} {\bibinfo {author} {\bibfnamefont {M.}~\bibnamefont {Stefaniak}},\ }\bibfield  {title} {\bibinfo {title} {{Proton-cluster femtoscopy with the HADES experiment}},\ }\href {https://doi.org/10.1051/epjconf/202429602001} {\bibfield  {journal} {\bibinfo  {journal} {EPJ Web Conf.}\ }\textbf {\bibinfo {volume} {296}},\ \bibinfo {pages} {02001} (\bibinfo {year} {2024}{\natexlab{b}})},\ \Eprint {https://arxiv.org/abs/2401.12966} {arXiv:2401.12966 [nucl-ex]} \BibitemShut {NoStop}%
\bibitem [{\citenamefont {Adamczyk}\ \emph {et~al.}(2015)\citenamefont {Adamczyk} \emph {et~al.}}]{STAR:2014shf}%
  \BibitemOpen
  \bibfield  {author} {\bibinfo {author} {\bibfnamefont {L.}~\bibnamefont {Adamczyk}} \emph {et~al.} (\bibinfo {collaboration} {STAR}),\ }\bibfield  {title} {\bibinfo {title} {{Beam-energy-dependent two-pion interferometry and the freeze-out eccentricity of pions measured in heavy ion collisions at the STAR detector}},\ }\href {https://doi.org/10.1103/PhysRevC.92.014904} {\bibfield  {journal} {\bibinfo  {journal} {Phys. Rev. C}\ }\textbf {\bibinfo {volume} {92}},\ \bibinfo {pages} {014904} (\bibinfo {year} {2015})},\ \Eprint {https://arxiv.org/abs/1403.4972} {arXiv:1403.4972 [nucl-ex]} \BibitemShut {NoStop}%
\bibitem [{\citenamefont {Aamodt}\ \emph {et~al.}(2011)\citenamefont {Aamodt} \emph {et~al.}}]{ALICE:2011dyt}%
  \BibitemOpen
  \bibfield  {author} {\bibinfo {author} {\bibfnamefont {K.}~\bibnamefont {Aamodt}} \emph {et~al.} (\bibinfo {collaboration} {ALICE}),\ }\bibfield  {title} {\bibinfo {title} {{Two-pion Bose-Einstein correlations in central Pb-Pb collisions at $\sqrt{{s}_{NN}} =$ 2.76 TeV}},\ }\href {https://doi.org/10.1016/j.physletb.2010.12.053} {\bibfield  {journal} {\bibinfo  {journal} {Phys. Lett. B}\ }\textbf {\bibinfo {volume} {696}},\ \bibinfo {pages} {328} (\bibinfo {year} {2011})},\ \Eprint {https://arxiv.org/abs/1012.4035} {arXiv:1012.4035 [nucl-ex]} \BibitemShut {NoStop}%
\bibitem [{\citenamefont {Adamczewski-Musch}\ \emph {et~al.}(2019)\citenamefont {Adamczewski-Musch} \emph {et~al.}}]{HADES:2018gop}%
  \BibitemOpen
  \bibfield  {author} {\bibinfo {author} {\bibfnamefont {J.}~\bibnamefont {Adamczewski-Musch}} \emph {et~al.} (\bibinfo {collaboration} {HADES}),\ }\bibfield  {title} {\bibinfo {title} {{Identical pion intensity interferometry in central Au + Au collisions at 1.23 A GeV}},\ }\href {https://doi.org/10.1016/j.physletb.2019.06.047} {\bibfield  {journal} {\bibinfo  {journal} {Phys. Lett. B}\ }\textbf {\bibinfo {volume} {795}},\ \bibinfo {pages} {446} (\bibinfo {year} {2019})},\ \Eprint {https://arxiv.org/abs/1811.06213} {arXiv:1811.06213 [nucl-ex]} \BibitemShut {NoStop}%
\bibitem [{\citenamefont {Voloshin}\ \emph {et~al.}(1997)\citenamefont {Voloshin}, \citenamefont {Lednicky}, \citenamefont {Panitkin},\ and\ \citenamefont {Xu}}]{Voloshin:1997jh}%
  \BibitemOpen
  \bibfield  {author} {\bibinfo {author} {\bibfnamefont {S.}~\bibnamefont {Voloshin}}, \bibinfo {author} {\bibfnamefont {R.}~\bibnamefont {Lednicky}}, \bibinfo {author} {\bibfnamefont {S.}~\bibnamefont {Panitkin}},\ and\ \bibinfo {author} {\bibfnamefont {N.}~\bibnamefont {Xu}},\ }\bibfield  {title} {\bibinfo {title} {{Relative space-time asymmetries in pion and nucleon production in noncentral nucleus-nucleus collisions at high-energies}},\ }\href {https://doi.org/10.1103/PhysRevLett.79.4766} {\bibfield  {journal} {\bibinfo  {journal} {Phys. Rev. Lett.}\ }\textbf {\bibinfo {volume} {79}},\ \bibinfo {pages} {4766} (\bibinfo {year} {1997})},\ \Eprint {https://arxiv.org/abs/nucl-th/9708044} {arXiv:nucl-th/9708044} \BibitemShut {NoStop}%
\bibitem [{\citenamefont {Zawisza}(2011{\natexlab{a}})}]{zawisza2011meson}%
  \BibitemOpen
  \bibfield  {author} {\bibinfo {author} {\bibfnamefont {M.}~\bibnamefont {Zawisza}},\ }\bibfield  {title} {\bibinfo {title} {Meson-baryon femtoscopy in au+ au collisions at 200 gev measured by star experiment},\ }\href@noop {} {\bibfield  {journal} {\bibinfo  {journal} {Indian Journal of Physics}\ }\textbf {\bibinfo {volume} {85}},\ \bibinfo {pages} {1051} (\bibinfo {year} {2011}{\natexlab{a}})}\BibitemShut {NoStop}%
\bibitem [{\citenamefont {Zawisza}(2011{\natexlab{b}})}]{Zawisza:2010az}%
  \BibitemOpen
  \bibfield  {author} {\bibinfo {author} {\bibfnamefont {M.}~\bibnamefont {Zawisza}} (\bibinfo {collaboration} {STAR}),\ }\bibfield  {title} {\bibinfo {title} {{Pion-proton correlations and asymmetry measurement in Au+Au collisions at $\sqrt{s_{NN}}=200$ $GeV$ data}},\ }\href {https://doi.org/10.1134/S1547477111090378} {\bibfield  {journal} {\bibinfo  {journal} {Phys. Part. Nucl. Lett.}\ }\textbf {\bibinfo {volume} {8}},\ \bibinfo {pages} {924} (\bibinfo {year} {2011}{\natexlab{b}})},\ \Eprint {https://arxiv.org/abs/1012.5666} {arXiv:1012.5666 [nucl-ex]} \BibitemShut {NoStop}%
\bibitem [{\citenamefont {Poniatowska}(2015)}]{Poniatowska:2015zca}%
  \BibitemOpen
  \bibfield  {author} {\bibinfo {author} {\bibfnamefont {K.}~\bibnamefont {Poniatowska}} (\bibinfo {collaboration} {STAR}),\ }\bibfield  {title} {\bibinfo {title} {{Pion-kaon femtoscopy in Au+Au collisions at STAR}},\ }\href {https://doi.org/10.1051/epjconf/20149504051} {\bibfield  {journal} {\bibinfo  {journal} {EPJ Web Conf.}\ }\textbf {\bibinfo {volume} {95}},\ \bibinfo {pages} {04051} (\bibinfo {year} {2015})}\BibitemShut {NoStop}%
\bibitem [{\citenamefont {Acharya}\ \emph {et~al.}(2021)\citenamefont {Acharya} \emph {et~al.}}]{ALICE:2020mkb}%
  \BibitemOpen
  \bibfield  {author} {\bibinfo {author} {\bibfnamefont {S.}~\bibnamefont {Acharya}} \emph {et~al.} (\bibinfo {collaboration} {ALICE}),\ }\bibfield  {title} {\bibinfo {title} {{Pion-kaon femtoscopy and the lifetime of the hadronic phase in Pb$-$Pb collisions at $\sqrt{s_{\rm{NN}}}$ = 2.76 TeV}},\ }\href {https://doi.org/10.1016/j.physletb.2020.136030} {\bibfield  {journal} {\bibinfo  {journal} {Phys. Lett. B}\ }\textbf {\bibinfo {volume} {813}},\ \bibinfo {pages} {136030} (\bibinfo {year} {2021})},\ \Eprint {https://arxiv.org/abs/2007.08315} {arXiv:2007.08315 [nucl-ex]} \BibitemShut {NoStop}%
\bibitem [{\citenamefont {Wang}\ \emph {et~al.}(2024)\citenamefont {Wang}, \citenamefont {Ma},\ and\ \citenamefont {Zhang}}]{Wang:2024ykl}%
  \BibitemOpen
  \bibfield  {author} {\bibinfo {author} {\bibfnamefont {T.-T.}\ \bibnamefont {Wang}}, \bibinfo {author} {\bibfnamefont {Y.-G.}\ \bibnamefont {Ma}},\ and\ \bibinfo {author} {\bibfnamefont {S.}~\bibnamefont {Zhang}},\ }\bibfield  {title} {\bibinfo {title} {Femtoscopy between $\pi$, $k$ and $p$ in different heavy-ion collisions at $\sqrt{s_{NN}} = 39$ gev},\ }\href@noop {} {\bibfield  {journal} {\bibinfo  {journal} {arXiv preprint arXiv:2401.12257}\ } (\bibinfo {year} {2024})}\BibitemShut {NoStop}%
\bibitem [{\citenamefont {Brown}\ and\ \citenamefont {Danielewicz}(1997)}]{Brown:1997ku}%
  \BibitemOpen
  \bibfield  {author} {\bibinfo {author} {\bibfnamefont {D.~A.}\ \bibnamefont {Brown}}\ and\ \bibinfo {author} {\bibfnamefont {P.}~\bibnamefont {Danielewicz}},\ }\bibfield  {title} {\bibinfo {title} {{Imaging of sources in heavy ion reactions}},\ }\href {https://doi.org/10.1016/S0370-2693(97)00251-7} {\bibfield  {journal} {\bibinfo  {journal} {Phys. Lett. B}\ }\textbf {\bibinfo {volume} {398}},\ \bibinfo {pages} {252} (\bibinfo {year} {1997})},\ \Eprint {https://arxiv.org/abs/nucl-th/9701010} {arXiv:nucl-th/9701010} \BibitemShut {NoStop}%
\bibitem [{\citenamefont {Bass}\ \emph {et~al.}(1998)\citenamefont {Bass} \emph {et~al.}}]{Bass:1998ca}%
  \BibitemOpen
  \bibfield  {author} {\bibinfo {author} {\bibfnamefont {S.~A.}\ \bibnamefont {Bass}} \emph {et~al.},\ }\bibfield  {title} {\bibinfo {title} {{Microscopic models for ultrarelativistic heavy ion collisions}},\ }\href {https://doi.org/10.1016/S0146-6410(98)00058-1} {\bibfield  {journal} {\bibinfo  {journal} {Prog. Part. Nucl. Phys.}\ }\textbf {\bibinfo {volume} {41}},\ \bibinfo {pages} {255} (\bibinfo {year} {1998})},\ \Eprint {https://arxiv.org/abs/nucl-th/9803035} {arXiv:nucl-th/9803035} \BibitemShut {NoStop}%
\bibitem [{\citenamefont {Bleicher}\ \emph {et~al.}(1999)\citenamefont {Bleicher} \emph {et~al.}}]{Bleicher:1999xi}%
  \BibitemOpen
  \bibfield  {author} {\bibinfo {author} {\bibfnamefont {M.}~\bibnamefont {Bleicher}} \emph {et~al.},\ }\bibfield  {title} {\bibinfo {title} {{Relativistic hadron hadron collisions in the ultrarelativistic quantum molecular dynamics model}},\ }\href {https://doi.org/10.1088/0954-3899/25/9/308} {\bibfield  {journal} {\bibinfo  {journal} {J. Phys. G}\ }\textbf {\bibinfo {volume} {25}},\ \bibinfo {pages} {1859} (\bibinfo {year} {1999})},\ \Eprint {https://arxiv.org/abs/hep-ph/9909407} {arXiv:hep-ph/9909407} \BibitemShut {NoStop}%
\bibitem [{\citenamefont {Bleicher}\ and\ \citenamefont {Bratkovskaya}(2022)}]{Bleicher:2022kcu}%
  \BibitemOpen
  \bibfield  {author} {\bibinfo {author} {\bibfnamefont {M.}~\bibnamefont {Bleicher}}\ and\ \bibinfo {author} {\bibfnamefont {E.}~\bibnamefont {Bratkovskaya}},\ }\bibfield  {title} {\bibinfo {title} {{Modelling relativistic heavy-ion collisions with dynamical transport approaches}},\ }\href {https://doi.org/10.1016/j.ppnp.2021.103920} {\bibfield  {journal} {\bibinfo  {journal} {Prog. Part. Nucl. Phys.}\ }\textbf {\bibinfo {volume} {122}},\ \bibinfo {pages} {103920} (\bibinfo {year} {2022})}\BibitemShut {NoStop}%
\bibitem [{\citenamefont {Omana~Kuttan}\ \emph {et~al.}(2022)\citenamefont {Omana~Kuttan}, \citenamefont {Motornenko}, \citenamefont {Steinheimer}, \citenamefont {Stoecker}, \citenamefont {Nara},\ and\ \citenamefont {Bleicher}}]{OmanaKuttan:2022the}%
  \BibitemOpen
  \bibfield  {author} {\bibinfo {author} {\bibfnamefont {M.}~\bibnamefont {Omana~Kuttan}}, \bibinfo {author} {\bibfnamefont {A.}~\bibnamefont {Motornenko}}, \bibinfo {author} {\bibfnamefont {J.}~\bibnamefont {Steinheimer}}, \bibinfo {author} {\bibfnamefont {H.}~\bibnamefont {Stoecker}}, \bibinfo {author} {\bibfnamefont {Y.}~\bibnamefont {Nara}},\ and\ \bibinfo {author} {\bibfnamefont {M.}~\bibnamefont {Bleicher}},\ }\bibfield  {title} {\bibinfo {title} {{A chiral mean-field equation-of-state in UrQMD: effects on the heavy ion compression stage}},\ }\href {https://doi.org/10.1140/epjc/s10052-022-10400-2} {\bibfield  {journal} {\bibinfo  {journal} {Eur. Phys. J. C}\ }\textbf {\bibinfo {volume} {82}},\ \bibinfo {pages} {427} (\bibinfo {year} {2022})},\ \Eprint {https://arxiv.org/abs/2201.01622} {arXiv:2201.01622 [nucl-th]} \BibitemShut {NoStop}%
\bibitem [{\citenamefont {Steinheimer}\ \emph {et~al.}(2022)\citenamefont {Steinheimer}, \citenamefont {Motornenko}, \citenamefont {Sorensen}, \citenamefont {Nara}, \citenamefont {Koch},\ and\ \citenamefont {Bleicher}}]{Steinheimer:2022gqb}%
  \BibitemOpen
  \bibfield  {author} {\bibinfo {author} {\bibfnamefont {J.}~\bibnamefont {Steinheimer}}, \bibinfo {author} {\bibfnamefont {A.}~\bibnamefont {Motornenko}}, \bibinfo {author} {\bibfnamefont {A.}~\bibnamefont {Sorensen}}, \bibinfo {author} {\bibfnamefont {Y.}~\bibnamefont {Nara}}, \bibinfo {author} {\bibfnamefont {V.}~\bibnamefont {Koch}},\ and\ \bibinfo {author} {\bibfnamefont {M.}~\bibnamefont {Bleicher}},\ }\bibfield  {title} {\bibinfo {title} {{The high-density equation of state in heavy-ion collisions: constraints from proton flow}},\ }\href {https://doi.org/10.1140/epjc/s10052-022-10894-w} {\bibfield  {journal} {\bibinfo  {journal} {Eur. Phys. J. C}\ }\textbf {\bibinfo {volume} {82}},\ \bibinfo {pages} {911} (\bibinfo {year} {2022})},\ \Eprint {https://arxiv.org/abs/2208.12091} {arXiv:2208.12091 [nucl-th]} \BibitemShut {NoStop}%
\bibitem [{\citenamefont {Savchuk}\ \emph {et~al.}(2023{\natexlab{a}})\citenamefont {Savchuk}, \citenamefont {Poberezhnyuk}, \citenamefont {Motornenko}, \citenamefont {Steinheimer}, \citenamefont {Gorenstein},\ and\ \citenamefont {Vovchenko}}]{Savchuk:2022msa}%
  \BibitemOpen
  \bibfield  {author} {\bibinfo {author} {\bibfnamefont {O.}~\bibnamefont {Savchuk}}, \bibinfo {author} {\bibfnamefont {R.~V.}\ \bibnamefont {Poberezhnyuk}}, \bibinfo {author} {\bibfnamefont {A.}~\bibnamefont {Motornenko}}, \bibinfo {author} {\bibfnamefont {J.}~\bibnamefont {Steinheimer}}, \bibinfo {author} {\bibfnamefont {M.~I.}\ \bibnamefont {Gorenstein}},\ and\ \bibinfo {author} {\bibfnamefont {V.}~\bibnamefont {Vovchenko}},\ }\bibfield  {title} {\bibinfo {title} {{Phase transition amplification of proton number fluctuations in nuclear collisions from a transport model approach}},\ }\href {https://doi.org/10.1103/PhysRevC.107.024913} {\bibfield  {journal} {\bibinfo  {journal} {Phys. Rev. C}\ }\textbf {\bibinfo {volume} {107}},\ \bibinfo {pages} {024913} (\bibinfo {year} {2023}{\natexlab{a}})},\ \Eprint {https://arxiv.org/abs/2211.13200} {arXiv:2211.13200 [hep-ph]} \BibitemShut {NoStop}%
\bibitem [{\citenamefont {Steinheimer}\ \emph {et~al.}(2011{\natexlab{a}})\citenamefont {Steinheimer}, \citenamefont {Schramm},\ and\ \citenamefont {Stocker}}]{Steinheimer:2010ib}%
  \BibitemOpen
  \bibfield  {author} {\bibinfo {author} {\bibfnamefont {J.}~\bibnamefont {Steinheimer}}, \bibinfo {author} {\bibfnamefont {S.}~\bibnamefont {Schramm}},\ and\ \bibinfo {author} {\bibfnamefont {H.}~\bibnamefont {Stocker}},\ }\bibfield  {title} {\bibinfo {title} {{An Effective chiral Hadron-Quark Equation of State}},\ }\href {https://doi.org/10.1088/0954-3899/38/3/035001} {\bibfield  {journal} {\bibinfo  {journal} {J. Phys. G}\ }\textbf {\bibinfo {volume} {38}},\ \bibinfo {pages} {035001} (\bibinfo {year} {2011}{\natexlab{a}})},\ \Eprint {https://arxiv.org/abs/1009.5239} {arXiv:1009.5239 [hep-ph]} \BibitemShut {NoStop}%
\bibitem [{\citenamefont {Steinheimer}\ \emph {et~al.}(2011{\natexlab{b}})\citenamefont {Steinheimer}, \citenamefont {Schramm},\ and\ \citenamefont {Stocker}}]{Steinheimer:2011ea}%
  \BibitemOpen
  \bibfield  {author} {\bibinfo {author} {\bibfnamefont {J.}~\bibnamefont {Steinheimer}}, \bibinfo {author} {\bibfnamefont {S.}~\bibnamefont {Schramm}},\ and\ \bibinfo {author} {\bibfnamefont {H.}~\bibnamefont {Stocker}},\ }\bibfield  {title} {\bibinfo {title} {{The hadronic SU(3) Parity Doublet Model for Dense Matter, its extension to quarks and the strange equation of state}},\ }\href {https://doi.org/10.1103/PhysRevC.84.045208} {\bibfield  {journal} {\bibinfo  {journal} {Phys. Rev. C}\ }\textbf {\bibinfo {volume} {84}},\ \bibinfo {pages} {045208} (\bibinfo {year} {2011}{\natexlab{b}})},\ \Eprint {https://arxiv.org/abs/1108.2596} {arXiv:1108.2596 [hep-ph]} \BibitemShut {NoStop}%
\bibitem [{\citenamefont {Mukherjee}\ \emph {et~al.}(2017)\citenamefont {Mukherjee}, \citenamefont {Steinheimer},\ and\ \citenamefont {Schramm}}]{Mukherjee:2016nhb}%
  \BibitemOpen
  \bibfield  {author} {\bibinfo {author} {\bibfnamefont {A.}~\bibnamefont {Mukherjee}}, \bibinfo {author} {\bibfnamefont {J.}~\bibnamefont {Steinheimer}},\ and\ \bibinfo {author} {\bibfnamefont {S.}~\bibnamefont {Schramm}},\ }\bibfield  {title} {\bibinfo {title} {{Higher-order baryon number susceptibilities: interplay between the chiral and the nuclear liquid-gas transitions}},\ }\href {https://doi.org/10.1103/PhysRevC.96.025205} {\bibfield  {journal} {\bibinfo  {journal} {Phys. Rev. C}\ }\textbf {\bibinfo {volume} {96}},\ \bibinfo {pages} {025205} (\bibinfo {year} {2017})},\ \Eprint {https://arxiv.org/abs/1611.10144} {arXiv:1611.10144 [nucl-th]} \BibitemShut {NoStop}%
\bibitem [{\citenamefont {Motornenko}\ \emph {et~al.}(2019)\citenamefont {Motornenko}, \citenamefont {Vovchenko}, \citenamefont {Steinheimer}, \citenamefont {Schramm},\ and\ \citenamefont {Stoecker}}]{Motornenko:2018hjw}%
  \BibitemOpen
  \bibfield  {author} {\bibinfo {author} {\bibfnamefont {A.}~\bibnamefont {Motornenko}}, \bibinfo {author} {\bibfnamefont {V.}~\bibnamefont {Vovchenko}}, \bibinfo {author} {\bibfnamefont {J.}~\bibnamefont {Steinheimer}}, \bibinfo {author} {\bibfnamefont {S.}~\bibnamefont {Schramm}},\ and\ \bibinfo {author} {\bibfnamefont {H.}~\bibnamefont {Stoecker}},\ }\bibfield  {title} {\bibinfo {title} {{QCD at high density: Equation of state for nuclear collisions and neutron stars}},\ }\href {https://doi.org/10.1016/j.nuclphysa.2018.11.028} {\bibfield  {journal} {\bibinfo  {journal} {Nucl. Phys. A}\ }\textbf {\bibinfo {volume} {982}},\ \bibinfo {pages} {891} (\bibinfo {year} {2019})},\ \Eprint {https://arxiv.org/abs/1809.02000} {arXiv:1809.02000 [hep-ph]} \BibitemShut {NoStop}%
\bibitem [{\citenamefont {Motornenko}\ \emph {et~al.}(2020)\citenamefont {Motornenko}, \citenamefont {Steinheimer}, \citenamefont {Vovchenko}, \citenamefont {Schramm},\ and\ \citenamefont {Stoecker}}]{Motornenko:2019arp}%
  \BibitemOpen
  \bibfield  {author} {\bibinfo {author} {\bibfnamefont {A.}~\bibnamefont {Motornenko}}, \bibinfo {author} {\bibfnamefont {J.}~\bibnamefont {Steinheimer}}, \bibinfo {author} {\bibfnamefont {V.}~\bibnamefont {Vovchenko}}, \bibinfo {author} {\bibfnamefont {S.}~\bibnamefont {Schramm}},\ and\ \bibinfo {author} {\bibfnamefont {H.}~\bibnamefont {Stoecker}},\ }\bibfield  {title} {\bibinfo {title} {{Equation of state for hot QCD and compact stars from a mean field approach}},\ }\href {https://doi.org/10.1103/PhysRevC.101.034904} {\bibfield  {journal} {\bibinfo  {journal} {Phys. Rev. C}\ }\textbf {\bibinfo {volume} {101}},\ \bibinfo {pages} {034904} (\bibinfo {year} {2020})},\ \Eprint {https://arxiv.org/abs/1905.00866} {arXiv:1905.00866 [hep-ph]} \BibitemShut {NoStop}%
\bibitem [{\citenamefont {Adamczewski-Musch}\ \emph {et~al.}(2018)\citenamefont {Adamczewski-Musch} \emph {et~al.}}]{HADES:2017def}%
  \BibitemOpen
  \bibfield  {author} {\bibinfo {author} {\bibfnamefont {J.}~\bibnamefont {Adamczewski-Musch}} \emph {et~al.} (\bibinfo {collaboration} {HADES}),\ }\bibfield  {title} {\bibinfo {title} {{Centrality determination of Au + Au collisions at 1.23A GeV with HADES}},\ }\href {https://doi.org/10.1140/epja/i2018-12513-7} {\bibfield  {journal} {\bibinfo  {journal} {Eur. Phys. J. A}\ }\textbf {\bibinfo {volume} {54}},\ \bibinfo {pages} {85} (\bibinfo {year} {2018})},\ \Eprint {https://arxiv.org/abs/1712.07993} {arXiv:1712.07993 [nucl-ex]} \BibitemShut {NoStop}%
\bibitem [{\citenamefont {Pratt}()}]{coral}%
  \BibitemOpen
  \bibfield  {author} {\bibinfo {author} {\bibfnamefont {S.}~\bibnamefont {Pratt}},\ }\href {https://github.com/scottedwardpratt/coral} {\bibinfo {title} {Correlations analysis library (coral)}}\BibitemShut {NoStop}%
\bibitem [{\citenamefont {Savchuk}\ \emph {et~al.}(2023{\natexlab{b}})\citenamefont {Savchuk}, \citenamefont {Motornenko}, \citenamefont {Steinheimer}, \citenamefont {Vovchenko}, \citenamefont {Bleicher}, \citenamefont {Gorenstein},\ and\ \citenamefont {Galatyuk}}]{Savchuk:2022aev}%
  \BibitemOpen
  \bibfield  {author} {\bibinfo {author} {\bibfnamefont {O.}~\bibnamefont {Savchuk}}, \bibinfo {author} {\bibfnamefont {A.}~\bibnamefont {Motornenko}}, \bibinfo {author} {\bibfnamefont {J.}~\bibnamefont {Steinheimer}}, \bibinfo {author} {\bibfnamefont {V.}~\bibnamefont {Vovchenko}}, \bibinfo {author} {\bibfnamefont {M.}~\bibnamefont {Bleicher}}, \bibinfo {author} {\bibfnamefont {M.}~\bibnamefont {Gorenstein}},\ and\ \bibinfo {author} {\bibfnamefont {T.}~\bibnamefont {Galatyuk}},\ }\bibfield  {title} {\bibinfo {title} {{Enhanced dilepton emission from a phase transition in dense matter}},\ }\href {https://doi.org/10.1088/1361-6471/acfccf} {\bibfield  {journal} {\bibinfo  {journal} {J. Phys. G}\ }\textbf {\bibinfo {volume} {50}},\ \bibinfo {pages} {125104} (\bibinfo {year} {2023}{\natexlab{b}})},\ \Eprint {https://arxiv.org/abs/2209.05267} {arXiv:2209.05267 [nucl-th]} \BibitemShut {NoStop}%
\bibitem [{\citenamefont {Reichert}\ \emph {et~al.}(2023)\citenamefont {Reichert}, \citenamefont {Savchuk}, \citenamefont {Kittiratpattana}, \citenamefont {Li}, \citenamefont {Steinheimer}, \citenamefont {Gorenstein},\ and\ \citenamefont {Bleicher}}]{Reichert:2023eev}%
  \BibitemOpen
  \bibfield  {author} {\bibinfo {author} {\bibfnamefont {T.}~\bibnamefont {Reichert}}, \bibinfo {author} {\bibfnamefont {O.}~\bibnamefont {Savchuk}}, \bibinfo {author} {\bibfnamefont {A.}~\bibnamefont {Kittiratpattana}}, \bibinfo {author} {\bibfnamefont {P.}~\bibnamefont {Li}}, \bibinfo {author} {\bibfnamefont {J.}~\bibnamefont {Steinheimer}}, \bibinfo {author} {\bibfnamefont {M.}~\bibnamefont {Gorenstein}},\ and\ \bibinfo {author} {\bibfnamefont {M.}~\bibnamefont {Bleicher}},\ }\bibfield  {title} {\bibinfo {title} {{Decoding the flow evolution in Au+Au reactions at 1.23A GeV using hadron flow correlations and dileptons}},\ }\href {https://doi.org/10.1016/j.physletb.2023.137947} {\bibfield  {journal} {\bibinfo  {journal} {Phys. Lett. B}\ }\textbf {\bibinfo {volume} {841}},\ \bibinfo {pages} {137947} (\bibinfo {year} {2023})},\ \Eprint {https://arxiv.org/abs/2302.13919} {arXiv:2302.13919 [nucl-th]} \BibitemShut {NoStop}%
\bibitem [{\citenamefont {Danielewicz}(1995)}]{Danielewicz:1994nb}%
  \BibitemOpen
  \bibfield  {author} {\bibinfo {author} {\bibfnamefont {P.}~\bibnamefont {Danielewicz}},\ }\bibfield  {title} {\bibinfo {title} {{Effects of compression and collective expansion on particle emission from central heavy ion reactions}},\ }\href {https://doi.org/10.1103/PhysRevC.51.716} {\bibfield  {journal} {\bibinfo  {journal} {Phys. Rev. C}\ }\textbf {\bibinfo {volume} {51}},\ \bibinfo {pages} {716} (\bibinfo {year} {1995})},\ \Eprint {https://arxiv.org/abs/nucl-th/9408018} {arXiv:nucl-th/9408018} \BibitemShut {NoStop}%
\bibitem [{\citenamefont {Weil}\ \emph {et~al.}(2016)\citenamefont {Weil} \emph {et~al.}}]{SMASH:2016zqf}%
  \BibitemOpen
  \bibfield  {author} {\bibinfo {author} {\bibfnamefont {J.}~\bibnamefont {Weil}} \emph {et~al.} (\bibinfo {collaboration} {SMASH}),\ }\bibfield  {title} {\bibinfo {title} {{Particle production and equilibrium properties within a new hadron transport approach for heavy-ion collisions}},\ }\href {https://doi.org/10.1103/PhysRevC.94.054905} {\bibfield  {journal} {\bibinfo  {journal} {Phys. Rev. C}\ }\textbf {\bibinfo {volume} {94}},\ \bibinfo {pages} {054905} (\bibinfo {year} {2016})},\ \Eprint {https://arxiv.org/abs/1606.06642} {arXiv:1606.06642 [nucl-th]} \BibitemShut {NoStop}%
\bibitem [{\citenamefont {Savchuk}(2025{\natexlab{b}})}]{Savchuk:2025prep}%
  \BibitemOpen
  \bibfield  {author} {\bibinfo {author} {\bibfnamefont {O.}~\bibnamefont {Savchuk}},\ }\bibfield  {title} {\bibinfo {title} {Manuscript in preparation}} (\bibinfo {year} {2025}{\natexlab{b}}),\ \bibinfo {note} {in preparation}\BibitemShut {NoStop}%
\bibitem [{\citenamefont {Nzabahimana}\ \emph {et~al.}(2025)\citenamefont {Nzabahimana}, \citenamefont {Danielewicz},\ and\ \citenamefont {Verde}}]{Nzabahimana:2025ivc}%
  \BibitemOpen
  \bibfield  {author} {\bibinfo {author} {\bibfnamefont {P.}~\bibnamefont {Nzabahimana}}, \bibinfo {author} {\bibfnamefont {P.}~\bibnamefont {Danielewicz}},\ and\ \bibinfo {author} {\bibfnamefont {G.}~\bibnamefont {Verde}},\ }\bibfield  {title} {\bibinfo {title} {Transport theory and correlation measurements: Coming to terms on emission sources},\ }\href@noop {} {\bibfield  {journal} {\bibinfo  {journal} {arXiv preprint arXiv:2506.01271}\ } (\bibinfo {year} {2025})}\BibitemShut {NoStop}%
\bibitem [{\citenamefont {Danielewicz}\ and\ \citenamefont {Pratt}(2007)}]{Danielewicz:2006hi}%
  \BibitemOpen
  \bibfield  {author} {\bibinfo {author} {\bibfnamefont {P.}~\bibnamefont {Danielewicz}}\ and\ \bibinfo {author} {\bibfnamefont {S.}~\bibnamefont {Pratt}},\ }\bibfield  {title} {\bibinfo {title} {{Analyzing correlation functions with tesseral and cartesian spherical harmonics}},\ }\href {https://doi.org/10.1103/PhysRevC.75.034907} {\bibfield  {journal} {\bibinfo  {journal} {Phys. Rev. C}\ }\textbf {\bibinfo {volume} {75}},\ \bibinfo {pages} {034907} (\bibinfo {year} {2007})},\ \Eprint {https://arxiv.org/abs/nucl-th/0612076} {arXiv:nucl-th/0612076} \BibitemShut {NoStop}%
\bibitem [{\citenamefont {Savchuk}(2025{\natexlab{c}})}]{Savchuk:2025prep2}%
  \BibitemOpen
  \bibfield  {author} {\bibinfo {author} {\bibfnamefont {O.}~\bibnamefont {Savchuk}},\ }\bibfield  {title} {\bibinfo {title} {Manuscript in preparation}} (\bibinfo {year} {2025}{\natexlab{c}}),\ \bibinfo {note} {in preparation}\BibitemShut {NoStop}%
\bibitem [{\citenamefont {Lisa}\ \emph {et~al.}(2021)\citenamefont {Lisa}, \citenamefont {Barbon}, \citenamefont {Chinellato}, \citenamefont {Serenone}, \citenamefont {Shen}, \citenamefont {Takahashi},\ and\ \citenamefont {Torrieri}}]{Lisa:2021zkj}%
  \BibitemOpen
  \bibfield  {author} {\bibinfo {author} {\bibfnamefont {M.~A.}\ \bibnamefont {Lisa}}, \bibinfo {author} {\bibfnamefont {J.~G.~P.}\ \bibnamefont {Barbon}}, \bibinfo {author} {\bibfnamefont {D.~D.}\ \bibnamefont {Chinellato}}, \bibinfo {author} {\bibfnamefont {W.~M.}\ \bibnamefont {Serenone}}, \bibinfo {author} {\bibfnamefont {C.}~\bibnamefont {Shen}}, \bibinfo {author} {\bibfnamefont {J.}~\bibnamefont {Takahashi}},\ and\ \bibinfo {author} {\bibfnamefont {G.}~\bibnamefont {Torrieri}},\ }\bibfield  {title} {\bibinfo {title} {{Vortex rings from high energy central p+A collisions}},\ }\href {https://doi.org/10.1103/PhysRevC.104.L011901} {\bibfield  {journal} {\bibinfo  {journal} {Phys. Rev. C}\ }\textbf {\bibinfo {volume} {104}},\ \bibinfo {pages} {011901} (\bibinfo {year} {2021})},\ \Eprint {https://arxiv.org/abs/2101.10872} {arXiv:2101.10872 [hep-ph]} \BibitemShut {NoStop}%
\end{thebibliography}%
\end{document}